\documentstyle[11pt,aaspp4]{article}

\def\deg{\ifmmode^\circ\ \else$^\circ$\ \fi}
\def\degcomma{$^\circ$,}
\def\degparen{$^\circ$)}
\def\degsemicol{$^\circ$;}
\def\degtimes{^\circ\times}
\def\ptdeg{\ifmmode\!^\circ\ \else$\!^\circ$\ \fi}

\def\fluenceunits{ergs-cm$^{-2}$}
\def\fluxunits{ergs-cm$^{-2}$-s$^{-1}$}
\def\cm2{cm$^2$}
\def\2sig{2$\sigma$}

\def\gray{$\gamma$--ray}
\def\grays{$\gamma$--rays}

\def\given{{~\vert~}}

\def\like{{\cal L}}

\def\bodds{{\lambda}}
\def\hyp{{\bf \cal H}}

\def\euler{{\rm e }}

\received{1997 November 11}

\lefthead{Connors and Hueter}
\righthead{X--ray Characteristics of a classic GRB}

\begin{document}

%
%
\title{
The X--ray Characteristics of a Classical Gamma--Ray Burst
and its Afterglow}

\author{A. Connors}
\affil{ISEOS, University of New Hampshire, Durham, NH 03824}
\authoraddr{Morse Hall, ISEOS, University of New Hampshire, Durham, NH 03824}
\and
\author{Geoffrey J. Hueter\altaffilmark{1}}
\affil{Center for Astrophysics and Space Sciences,
University of California, San Diego}
\altaffiltext{1}{Now at Far Point Technology, P.O. Box 421467, San Diego CA 92142}

\begin{abstract}
The serendipitous observation of GRB~780506
by co--aligned {\gray} ({\it HEAO~1} A--4 0.02 -- 6~MeV)
and X--ray ({\it HEAO~1} A--2 2--60~keV) instruments 
during a six hr pointing at a blank section of the sky
gave us unprecedented high signal--to--noise X--ray spectra
and light curves of a {\gray} burst and its afterglow.
First, the spectra:
although the time resolution was only 10.24 s, 
it was possible to derive unambiguous position constraints and
 a reliable instrument response.
We therefore could see two breaks in the initial spectrum, one 
consistent with a peak in $\nu F_{\nu}$ of $\sim$45 keV,
and one below 4 keV, consistent with strong absorption.
The event
exhibited dramatic spectral variability, hardening as it
rose to each peak, and softening as it fell, similar
to behavior reported above 100~keV from other bursts.
The initial strong turnover below a few keV evolved into a
slight excess as the burst progressed.
The spectral shape varied widely outside low energy limits
prescribed by current relativistic shock models.
At no time was there
evidence for any lines, either in emission or absorption.
Overall, $\sim$20\% of the total
energy was radiated in nonthermal 2--30~keV X--rays.
Second, the light curve:
the X--ray proportional counters
were able to monitor the flux for an hour before and some hours after
burst onset.
Two minutes after it ended, {\it HEAO~1} A--2 detected a faint resurgence
of 2--10~keV flux, rising to a peak $\sim$seven minutes after burst onset,
followed by irregular emission
with best--fit decay time of $\sim{{1}\over{2}}$ hr.
The position of the source of the extended flux is consistent with
that of the burst, and probably not
from a serendipitous transient.
We estimated that the entire afterglow
radiated  between 3 and 30\% of the $>$1~keV energy radiated
during the burst.
In this paper, we present the 2--300~keV {\it HEAO~1} A--2 and A--4
light curves and spectra from GRB~780506, 
and its apparent $\sim$1~keV afterglow.
\end{abstract}

\keywords
{X--rays: general --- gamma--rays: bursts --- methods: statistical}

\section{ Introduction}

We present a serendipitous observation
of 2--300~keV X--rays from a ``classic'' 
{\gray} burst, GRB~780506,
observed on May 6, 1978, during a $\sim$6 hr pointed maneuver.
The event was discovered in {\it HEAO~1}
A--4 (0.03--6~MeV) data (Hueter\markcite{hu1} 1987), and subsequently found to have
been detected by {\it HEAO~1} A--2 (2--60~keV).
The detectors observed
the same region of sky for about $1{{1}\over{2}}$
hr before and $4{{1}\over{2}}$ hr after the event.
We obtained complete coverage of the {\gray} burst with instruments
spanning three decades in energy;
and reasonably continuous coverage (the source was
eclipsed by the Earth about 30\% of the time) of the
associated fainter longer time-scale emission,
which we had termed the afterglow. 
(See Connors and McConnell\markcite{co96} 1996 for a preliminary report.)
Although the event was faint at higher energies,
($\sim 7\times10^{-7}$\fluenceunits ~0.03--6~MeV),
below 60~keV
the small field of view, low background, good position constraints,
and $\sim$400 cm$^2$ effective area 
combined to give high signal--to--noise data down to 2~keV.
We obtained good signal--to--noise 2--60~keV spectra every
10.24~s, with well--characterized instrument response.
This is a rare combination.

Until the launch of the wide field of view {\it Ginga}
satellite,
out of hundreds of {\gray} bursts, $<$10~keV X--rays
had been detected from only a handful (Wheaton et al.\markcite{wh1} 1973;
Cline et al.\markcite{cl1} 1979; Gilman et al.\markcite{gi1} 1980; 
Terrell et al.\markcite{te2} 1982; 
Laros et al.\markcite{la1}\markcite{la2} 1984a,b;  
Katoh et al.\markcite{ka2} 1984).
These earliest measurements of $<$10~keV X--rays from {\gray}
bursts suffered from small collecting areas
($\sim$several cm$^2$) and
ill-constrained positions
(Wheaton et al.\markcite{wh1} 1973;  Gilman et al.\markcite{gi1} 1980).
Many were limited to data from scanning instruments,
so that $<$10~keV light curves
were sampled in snapshots at regular intervals;
or did not have continuous spectral coverage from {\grays} down to X--rays
(Laros et al.\markcite{la1}\markcite{la2} 1984a,b, 
Katoh et al.\markcite{ka2} 1984; Terrell et al.\markcite{te2} 1982).
In the past decade, the large field of view $\gamma$--ray burst
detectors, on board the 
Japanese satellite {\it Ginga},
have been able to measure X--rays from a significant sample, 
observing both line features
(Murakami et al.\markcite{mu1} 1988; Fenimore et al.\markcite{fe1} 1988;
Yoshida et al.\markcite{yo2} 1992;  Graziani et al.\markcite{gr1} 1992),
and continuum characteristics (Strohmayer et al.\markcite{st1} 1998).
They observe a variety of $<$20~keV spectral shapes, from flattening
and occasional rollovers at a few keV, to continued increases
in roughly half their events.  However large field of view instruments
can bring with them problems of high background and poorly 
constrained positions.
For {\it Ginga}, below $\sim$10~keV, the 
spectra were uncertain by $30\%$ due to unknown incidence angles
in all but four of the events.
Also, the higher background inherent in a low spatial--resolution
X--ray instrument could preclude detecting any extended afterglow
unless it were exceptionally bright.

Recently, arc-minute localizations of nine bright {\gray}
bursts have been made available hours after burst onset,
due to the Italian--Dutch {\it BeppoSAX} satellite 
(Heise et al.\markcite{heise}\markcite{heise97b} 1997,1998;
Piro et al.\markcite{piro}\markcite{piro97b} 1997,1998a,b;
Costa et al.\markcite{co97a}\markcite{co97b} 1997,1998; 
Halpern et al.\markcite{halpern} 1997);
 {\it RXTE}'s All-Sky Monitor (Smith et al.\markcite{smith97} 1998); 
and coordinated efforts
 among {\it CGRO} and scans by {\it RXTE}'s PCA (Takeshima et al.\markcite{ta97} 1998
and references therein).
This has allowed more sensitive, small field of view instruments
to be repointed to search for X--ray and longer--wavelength
afterglows (for example Murakami et al.\markcite{mu97} 1998).  
In at least seven of the nine recent well--localized
events, faint ($<$few$\times10^{-12}$\fluxunits) fading afterglows
were detected, typically hours or days later.

By contrast, in these {\it HEAO~1} data we observe the burst position
directly after the event, allowing detection of the initial
resurgence in soft X--ray flux.  For this event, one sees this
soft emission peak about seven minutes after burst onset, or about
five minutes after flux from the initial event has returned 
to background.

The burst itself was softer than average, with a power--law photon index
of $-2.4$ above $\sim$45~keV, implying it may be a representative
of a soft subset of {\gray} burst emission 
(Pizzichini\markcite{pi95} 1995;  
Kouveliotou et al.\markcite{koetal96} 1996; 
Belli\markcite{be97} 1997; Pendleton et al.\markcite{pe97} 1997).
Nevertheless, many of the features observed in GRB~780506 were also visible,
with less statistical significance, in the earlier
measurements of X--rays from {\gray} bursts.
The soft X--rays in general lasted longer than
the higher energy emission; the event exhibited rapid spectral variability,
hardening as it rose to each peak and softening as it fell;
and there was a dearth of soft photons just prior to the first peak.
This continues a pattern often noted
in other bursts at higher energies:
the shape of the light curve at lower energy, in general,
follows that of the slightly higher energies,
but with a slight soft lag, and with a `smearing'
of the softer emission; and the emission softening from peak to peak
(Norris et al. 1986; Band et al. 1992; 1993; Ford et al. 1994;
Kargatis et al.\markcite{ka2} 1994).
Other observations have suggested these effects
become more pronounced below about 50 keV.
(For example see GRB~830801 in Laros and Nishimura\markcite{la3} 1986.)
Also, despite this event's relatively soft spectrum,
the X--ray spectra were too flat to be fit by any simple thermal model.
The $<$50~keV spectral shape evolved from a flat power--law 
(photon index $-1.1$) with significant absorption below a few keV,
to a steeper power--law with an increasing (but still small) amount
of excess emission below a few keV that could be modeled by a 
black--body component.
 In particular, $<$10\% could have been attributed to
cooling black-body emission
from reprocessing in optically thick material.
This suggests to us
that the bulk of X--rays were generated through the same non-thermal processes
driving the {\gray} spectral evolution,
rather than deriving from a separate thermal component,
in contrast to the conclusions of Laros and Nishimura (1986) and
Chernenko and Mitrofanov\markcite{ch1} (1994).

In the following section we describe the instrument, present the
GRB~780506 light-curves and spectra, and compare them with other
observations of X--rays from {$\gamma$--ray} bursts.  We conclude with
a brief summary in Section 3.
(Appendix~A details the Bayesian algorithms used to determine
the burst and afterglow positions.)

\section{Observations and Analysis}

\subsection{The Experiments}

The {\it HEAO~1} satellite observed the sky from August 1977 to January 1979
with four different sky survey experiments
(A--1, A--2, A--3, and A--4)
spanning the energy range 0.1 keV -- 6~MeV.
Only the two widest field--of--view instruments, 
A--4 (0.03 -- 10~MeV) and A--2 (0.1 -- 60~keV), 
detected GRB~780506.
The A--4 experiment consisted of
 three sets of sodium iodide/cesium iodide
phoswich detectors
(A--4 HED, 0.1--6~MeV 40\deg FWHM circular field of view;
four A--4 MEDs, 0.03--3~MeV, 16\deg FWHM circular fields of view; and
two A--4 LEDs, 10--200~keV, $1.7^{\circ}\times 20\deg$
 collimated slats; 
Matteson\markcite{ma1} 1978).
The six A--2 multiwire, multilayer proportional counters
were designed with overlapping medium and small fields of view ---
both within each detector and among detectors --
for stable and well--monitored internal and cosmic background
(Rothschild et al.\markcite{ro1} 1979).
We used these to constrain the position of the burst and afterglow.
The A--2 HED1 (xenon plus propane veto layer, 2--60~keV,
3$\degtimes$6\deg plus 3$\degtimes$3\deg )
look-direction was offset by 6\deg from the others
(A--2 MED, argon plus propane veto layer, 1--20~keV,
3$\degtimes$1.5\deg plus 3$\degtimes$3{\deg};
A--2 HED2, xenon, 1--60~keV,
3$\degtimes$6\deg plus 3$\degtimes$3{\deg}; and
A--2 HED3, xenon plus propane veto layer, 2--60~keV,
3$\degtimes$1.5\deg plus 3$\degtimes$3\deg ).
The A--2 LEDs
were turned off at this time.
GRB~780506 was detected only in the
A--4 MEDs and the offset A--2 HED1 detector.

\subsection{Burst detection}

The authors had independently performed statistical surveys of all
the {\it HEAO~1} A--4 data (10~keV--6~MeV; Hueter 1987) and the
``scanning'' portion of the {\it HEAO~1} A--2 data (2--60~keV; Connors
1988; Connors, Serlemitsos, and Swank 1986), for events with durations
comparable to those of {$\gamma$--ray} bursts.
Out of 21 {\gray} bursts uncovered in
the A--4 survey, only two were found in A--2 data with
greater than $\sim$3$\sigma$ significance.
The stronger of the two, found in A--2 scanning data,
was detected through the sides of the A--2 detectors
and so was not useful for observing $<$50~keV X--rays 
(Connors\markcite{co1} 1988).
The fainter of the two was found to have occurred during a 
 $\sim$6.5 hr pointed maneuver, in a region of sky containing
no previously known X--ray sources.
(Pointed data were not covered by the A--2 systematic survey of 
Connors\markcite{co1} 1988).
This event, occurring on May 6, 1978
and given the designation GRB~780506,
was the only {\gray} burst
observed through the aperture of an A--2 detector
(rather than through the sides).
Since the A--2 fields of view were roughly twenty times smaller
than the A--4 MED fields of view, this is about the rate
of A--2 detections one expects

Hueter\markcite{hu1} (1987) classified GRB~780506 as a faint
(fluence $\sim$7$\times$10$^{-7}$ \fluenceunits ~0.03--6~MeV) 
{\gray} burst,
with duration  $\sim$10 s, and a photon power law
spectral index (above 30~keV) of 2.4$\pm$0.3.
It was observed
with three of the A--4 MEDs (the fourth was covered
by the blocking crystal), with energy ranges 25--100~keV,
70--1900~keV, and 100--2900~keV.  The event was not
detected in either of the two A--4 LEDs, indicating that the
burst position was
outside their fields of view.
On the basis of its moderate duration, reasonably hard
spectrum, and spectral variablity,
we classified this event as a classic {\gray} burst, rather than
a soft {\gray} repeater (Norris et al.\markcite{no2} 1991).
In BATSE data it would have been classified as a member
of the longer subset of {\gray} bursts
(Kouveliotou et al.\markcite{ko1} 1994).

In contrast to the rates from the A--2 HED1 
offset xenon proportional counter,
which rose from a background rate of
$\sim$20 cts--s$^{-1}$ to over 500 cts--s$^{-1}$,
the rates from A--2 HED3,
an identical xenon proportional counter
with a slightly different look direction and field of view,
rose from 15 cts--s$^{-1}$ to 16 cts--s$^{-1}$
during the 10 s interval flagged by A--4 data as GRB~780506.
Therefore we concluded that A--2 HED1 had detected X--rays through
the front of its collimators,
rather than {\grays} or charged particles through its sides.

The burst occurred while the orientation of the satellite was held
fixed so that the A--1 instruments on the opposite
side of the spacecraft could point at
the bright X--ray binary Cygnus X-1.
During this maneuver,
the A--2 and A--4 experiments viewed a ``blank'' section
of the sky near the Galactic plane.
They viewed this same section of sky
three and six days after GRB~780506
as the A--1
experiment pointed at Cygnus X-1 several times.
Each ``blank sky point''
lasted for roughly 6 hr,
interrupted by 20--30 minutes of Earth eclipse
every $\sim$90 minute orbit.
The {\it HEAO~1} experiments also scanned this region
for several days before and after the event,
viewing the region for $\sim$1 minute every
35 minutes.

\placefigure{figure1}

In Figure~1 we present the A--2 HED1 Layer 1 2-20~keV light curve,
plotted in 122.88 s bins.
We subtracted a mean background rate determined from the first hour
of the blank sky point.
(The small bump visible in the first orbit is not significant,
as $\chi^2$ for a fit to a constant rate was 14. for 17 122.88~s bins,
or 354. for 362~5.12~s bins.)
To reduce times of high particle background, we have excluded
times with MacIlwain's $L$--parameter $>$1.4
(equivalent to times of low rigidity; Tennant\markcite{te1} 1983).
The burst itself is contained in the single highest $\sim$2 minute bin,
visible shortly after the second gap (due to Earth occultation).
One then observes a slow resurgence of 2--20~keV flux.
For A--2 HED1,  $\chi^2$ for a fit to a constant, zero, rate was 72.6 
for 21 122.88~s bins; or $\chi^2$ of 531. for 440 5.12~s bins,
 for the remainder of the orbit following the burst.
The count rate was constant in the 2--20~keV energy band of the A--2 HED3
(identical to A--2 HED1 save for look direction and field of view)
during this same time period 
($\chi^2$ of 13.8 per 21 122.88~s bins; or
 $\chi^2$ of 417. for 440 5.12~s bins
for a fit to a constant zero rate),
indicating again that the increased counts came from X--rays from the source
rather than charged particles
or {\grays} through the detector walls.
Also, the spectra of the high--flux points after the burst
(including those in the second half of that orbit)
were quite soft, while instrumental background has quite a hard spectrum.

Given the position error--box (see below),
and rates of serendipitous X--ray sources (Connors et al. 1986,
Motch et al. 1991a,b; Linsky 1990),
Connors and McConnell\markcite{co96} (1996) calculated a $>98.5\%$ probability
that the longer--duration emission came from the same source as the burst.
Additionally, using ROSAT data they searched 
for a serendipitous X--ray transient
that could have produced one or both, with null result.
We therefore had associated this soft, slowly varying emission with
the burster, and had termed it the afterglow (Connors and McConnell
1995,1996; Connors 1988).

\subsection{Position Constraints}

Only the A--2 offset detector, HED1, showed a significant increase
in counts at the time of the burst, constraining the source
to be more distant than $-$6\deg
(along the satellite spin--direction) from the main detector axis.
To further constrain the source position,
we used a Bayesian algorithm that compared the measured rates
in the overlapping wide and narrow fields of view,
every 5.12~s, against a model
position, background rates, and intensity in that time bin 
(Appendix~A).
The orbit preceding the burst was used to estimate the prior probability
of the background rates (see Figure 1).
This was done separately for the burst itself
and the extended emission following the burst (Appendix~A).
For GRB~780506,
we calculated the best fit position to be at 2000.0 R.A. and Dec.
$\sim$7$^h$42$^m$48$^s$, $-41\deg51\arcmin18\arcsec$
or Galactic l$^{II}$,b$^{II}$ of 255.6\degcomma $-$9.0\degcomma
with $3\sigma$ (99.73\%) 
credible limits of $\pm$0.16\deg along the satellite spin direction,
and $\pm$3\deg perpendicular to it.
Although not exactly a quadrilateral, 
the corners of this A--2 $3\sigma$ credible region were roughly
118.9\degcomma $-$42.73{\deg};  
112.5\degcomma $-$40.56{\deg};
118.9\degcomma $-$43.04{\deg};  and 
112.5\degcomma $-$40.87{\deg}
(2000.0 R.A. and Dec).
For the extended emission, the $3\sigma$ contours were approximately
centered on those of GRB~780506, with a width of $\sim\pm0.5\deg$.
The corners of its A--2 $3\sigma$ credible region were roughly
118.8\degcomma $-$42.36\degsemicol  113.2\degcomma $-$40.50\degsemicol
118.3\degcomma $-$43.31\degsemicol  and 112.7\degcomma $-$41.43\deg
(2000.0 R.A. and Dec).

To further constrain the source position in the direction perpendicular
to the satellite spin plane,
we compared rates from A--2 HED1 and the much wider field of view A--4 MEDs,
in the $\sim$30--60~keV range that their spectral windows overlapped.
We first calculated the incident A--2 and A--4 photons-cm$^{-2}$-s$^{-1}$
in the four overlapping energy channels,
assuming the best-fit spectrum discussed in \S 2.6.
(Since the GRB~780506 spectra were relatively smooth and featureless,
our results were insensitive to the specific model employed,
as long as the fits were reasonably good.)
A least $\chi^2$ fit minimizing the differences in incident A--2/A--4
photon flux
in 4 channels in the 30-60~keV region of overlap,
during the first, brightest, 10.24 s of the burst,
showed the source position to lie $\pm$1\deg
(with a 1$\sigma$ uncertainty of 0.5\degparen ~
from the satellite spin plane,
with A--2/A--4 best fit positions of
116.9\degcomma $-$42.26\degsemicol and 114.5\degcomma $-$41.42\deg
(2000.0 R.A. and Dec.), or about 7.3\deg off the main detector axis.
The {\gray} burster was located outside the A--3 field of view;
and although it appeared to be on the edge
of one module of the A--1 field of view,
exceptionally high
background rates at the time of the burst
precluded using the A--1 rates to further constrain the position
(J. Norris 1988, private communication).
For the A--2 burst light--curves (Figure 2) and spectral analysis
(Figures 3 and 4),
we arbitrarily assumed one of the two best fit positions given above
( 114.5\degcomma $-$41.42\deg 2000.0 R.A. and Dec.),
although this choice makes little difference to the instrument response.

\subsection{ Burst Light Curves and Hardness Ratios}

In Figure 2, we present
the A--2 and A--4 light--curves 
during just the single highest 122.8~s time bin of Figure 1.
We used time bins of 1.28 s,
which was the shortest integration time
available from the A--2 HED1 during this observation.
At the top we have plotted
the A--4 count-rates for A--4 MED1 (90--2900~keV) and 
all A--4 MEDs below 200~keV (25--200 keV) for this event.
Directly below
are the 8--45~keV light curve from A--2 HED1
and the 2--6~keV light curve from the
same instrument.
At the bottom, we have displayed
the 8--45~keV/2--6~keV hardness ratio.
(The event was too faint for us to form meaningful
hardness ratios $>$100~keV.)
Several features hinted at in earlier reports of X--ray
counterparts are immediately apparent.
Overall,
the event softens as it progresses.
The emission hardened as it rose to each peak,
and softened as it fell.
A second main peak,
which was not visible in the $>$30~keV light curves,
was apparent in the 8--40~keV band.
In the 2--6~keV X--rays, it was brighter than the first main peak.
The longer wavelengths appear to have a longer decay time
than the shorter wavelengths, and although emission at
all wavelengths appears to begin to rise at the same time,
the softer light curve appears to peak on the order of
1~s after the harder ones.
This is consistent with what was observed for many
(but not all) {\it Ginga} bursts
(Strohmayer et al.\markcite{st1} 1998; 
Murakami et al.\markcite{mu3} 1992;  Yoshida et al.\markcite{yo1} 1989).

\placefigure{figure2}

At higher energies,
the evolution of different energy light--curves has been described
in two parts:  an increasing width of each peak as energy decreases;
and a soft lag.
Using BATSE data, Fenimore et al.\markcite{fe2} (1995)
have parametrized the average peak width as a function of energy
with a power--law, 
$W_{ac}(E) ~=~ 17.43 E^{-0.43};$
where $W_{ac}(E)$ is the width (in seconds) of
an average autocorrelation function at energy $E$,
measured where the value of the average autocorrelation is at
${\rm e}^{-{{1}\over{2}}}$ the value at its peak.
We compare this with the width of two peaks measured at
2--6~keV and 8--45~keV, in Figure~2.
Rather than taking the 
autocorrelation, we simply measure the width of each pulse
at ${\rm e}^{-{{1}\over{2}}}$ of the peak value.
For the first peak, one finds widths of
$\sim 7\pm 1$~s and $\sim 10\pm 1$~s at 8--45~keV and 2--6~keV,
respectively.  
For the second peak, one measures
$\sim 8\pm 1$~s and $\sim 12\pm 1$~s,
for the same energies.
(This second peak was not detected in the A--4 $>25$~keV data.)
Fenimore et al.\markcite{fe2} (1995)
point out the widths in each energy band will be dominated by
the timescale of the lowest energy in each.
Using their formula to extrapolate to X--ray energies, one
predicts timescales of $\sim 7.1$~s and 13~s, at 8~keV and 2~keV,
respectively;
in agreement with the values measured from GRB~780506.
That is, the same mechanism appears to drive the widths of the pulses
from a few hundred keV down to $\sim$2~keV.
We suggest that it is this soft lag and longer decay time
that could have been interpreted,
when incompletely sampled
or when continuous spectral coverage was not available
in earlier measurements, as a soft, thermal, component which varied
independently of the $>$ 30~keV {\grays} 
(Laros and Nishimura\markcite{la3} 1986).

Although a lag of the peak of softer energies with respect
to harder ones has often been noted at higher energies,
(Chernenko et al.\markcite{chernenko98} 1998; 
Chipman\markcite{ch2} 1994; Band et al.\markcite{ba1} 1992;
Norris et al.\markcite{no1} 1986),
it is not as well quantified as a function of energy.
It may be related to
an evolving low energy cut--off
seen both here, in GRB~780506; and at higher energies
(Pendleton et al.\markcite{pe1} 1996).

Vietri (1997), discussing the rather similar light--curve
of GRB~970228, attributes the second peak to an afterglow
due to a reverse (external)
shock.  However we note:
1) under that definition, even longer BATSE bursts with multiple,
widely separated peaks 
--- such as the well--known bright burst GRB~910503, BATSE trigger 143 ---
would have been classified as having afterglows;
2) when the 2--60~keV X--ray spectra are looked at in detail,
no evidence for a different kind of emission in the second peak
is apparent;  and
3) the wide variety of burst light--curves
 seen both at wavelengths $>20$~keV (Fishman, J.G. et al.\markcite{fi1}
1994)
 and below
(see references in Katoh et al.\markcite{ka2} 1984;
Laros et al.\markcite{la2} 1984a;
Laros and Nishimura\markcite{la3} 1986;
Murakami et al.\markcite{mu1,mu2} 1988, 1991;
and Strohmayer et al.\markcite{st1} 1998);
 caution against making such an identification from one or two bursts.

\subsection{Burst Spectra}

At the time of the burst, pulse height analyzed data were
read out every 10.24 s for A--2 HED1, and every 5.12 s for the
A--4 detectors.  We have marked the 10.24 s accumulation intervals
at the top of Figure 2 as intervals 1 through 6.
The A--4 instruments detected the burst only in the first
10.24 s, which contained the rise and
fall of the first peak.
In Figure 3, we display the $\nu F_{\nu}$ spectrum of the first peak
(intervals 1 and 2);  
second peak (intervals 3, 4, and 5); 
and the 600~s of the afterglow
during which it was brightest.
The photon--fluence and error bars of each point 
were determined from the best--fit model,
by scaling the observed counts and errors by the ratio of
the model to the predicted counts in each channel.
These were multiplied by the average energy in each bin
to obtain $\nu F_{\nu}$ spectra.
The A--2 and A--4 data were fit separately.
The A--4 data were well-fit by a simple power--law.
For the A--2 $<60$~keV spectra,
the only model that fit not only
the hard emission above 10~keV, but also
the initial strong low--energy turnover
and the growing soft excess below $\sim$6~keV
was a superposition of a power
law plus a black body, of the form
\begin{equation}
{{{\rm d}N}\over{{\rm d}E}} ~=~
\Bigl(
       A_1 {{ E^2 }\over{ \exp( E / kT ) - 1}} +
       A_2 E^{-\alpha}
\Bigr) {\rm e}^{-\sigma_{abs}N_H}
\end{equation}
where $\sigma_{abs} N_H$ is the amount of photoelectric absorption 
of a hydrogen column density $N_H$ at energy E
(assuming Galactic abundances).
(Other models we tried that did not fit included linear combinations of
power--laws,
thermal bremsstrahlung,  and thermal synchrotron spectra.
The frequently used ``GRB'' function of Band et al.\markcite{ba93} 1993 
does not model
an upturning spectrum and so was not a good representation
of the evolving $<$20~keV spectra.)
We emphasize that these were only
empirical fits, to illustrate the overall
shape of the spectrum in each interval.

\placefigure{figure3}

Overall,
we consider the X--ray to {\gray} spectra of {\gray} bursts
to consist of four parts:
1) the highest energy emission, which can extend in a power--law above
a break energy beyond tens of MeV to GeV energies
(Kippen et al.\markcite{ki1} 1998; Dingus et al.\markcite{di1} 1994; 
Matz et al.\markcite{ma2} 1985);
2) the intermediate energy emission, which tends to roll over
to a peak in $\nu F_{\nu}$ typically at $\sim$200~keV 
(but with a wide spread; Band et al.\markcite{ba2} 1993;
Kargatis et al.\markcite{ka1} 1994);  
3) the X--ray spectrum below 30--50~keV, which can either continue
or flatten even more
(Laros et al.\markcite{la1}\markcite{la2} 1984a,b; 
Strohmayer et al.\markcite{st1} 1998);
and 
4) a lower energy turnover, or flattening, or sometimes excess below 
several keV
(Strohmayer et al.\markcite{st1} 1998; Katoh et al.\markcite{ka2} 1984;
Preece et al.\markcite{pr1} 1996),
consistent with heavy intrinsic absorption, or excess thermal emission,
respectively.
One also typically (but not universally)
observes a softening trend in successive peaks
(Norris et al.\markcite{no1} 1986; Ford et al.\markcite{fo1} 1994; 
Band et al.\markcite{ba1} 1992).

From Figure~3, one sees GRB~780506 to have followed these patterns.
The bulk of the emission was emitted in the first peak.
The  $>$50 keV {\grays} of exhibited
a steeper than average spectrum
with a photon power law index of about $2.4\pm{0.3}.$
For all bursts in Hueter\markcite{hu1} (1987) the mean was 1.6$\pm$0.05.
This places GRB~780506 in the `soft' subclass of {\gray} bursts
suggested by Pizzichini\markcite{pi95} (1995); 
Belli\markcite{be97} (1997),
with a 100--300~keV / 50--100~keV fluence hardness ratio of $\sim$1.1
(while for an $E^{-1.6}$ spectrum it would be $\sim2.3$).
This subclass shows 
no evidence for turnover from a $\log(N>P) - \log(P)$ distribution
of $P^{-{{3}\over{2}}}$ 
(Belli\markcite{be97} 1997;
Kouveliotou et al.\markcite{koetal96} 1996; 
Pizzichini\markcite{pi95} 1995;  
Pendleton et al.\markcite{pe97} 1997).
For the first peak, 
the maximum in $\nu F_{\nu}$ occurred around 45~keV
where the A--2 and A--4 data overlap.
Below this the {\it HEAO~1} A--2 spectrum 
of the first peak is consistent with a power law
with photon index $1.17\pm{0.01}$.
There is an indication of a low energy turnover equivalent to
an absorption of 0.1$\pm$0.04$\times$10$^{23}$ cm$^{-2}$.

In these data, the distinction between the medium--energy turnover
($\sim45$~keV) and low--energy turnover (a few keV) is explicit.
This is not always the case in {\it Ginga} data 
(Strohmayer et al.\markcite{st1} 1998).

For the second peak, there is no significant A--4 emission
and the data are too uncertain to constrain a maximum in $\nu F_{\nu}$.
The A--2 emission is softer (photon index $1.8\pm{0.01}$)
and, instead of a turnover at low energies, shows emission
slightly in excess of the power law.
This we fit with a black-body spectrum
with temperature $1.8\pm{0.06}$~keV.

The spectrum of the afterglow was not well--constrained.
We found a $1.1\pm{0.4}$~keV black body with negligible absorption,
a $2.4{{+1.3}\atop {-0.8}}$ keV thin thermal bremsstrahlung model,
and a heavily absorbed, power law spectrum
with photon index of $3.9{{+1.8} \atop {-0.8}}$ all fit equally well.
(A power law with photon index $2.8\pm0.1$, with the absorption
fixed at zero, was also acceptable.)
We saw no evidence of hardening or softening during the afterglow.

However it is clear from Figure 2
that we have integrated over dramatic spectral variations.
Therefore in Figure 4.a--4.f
we display the spectra on shorter (10.24~s) timescales.
These we plotted in photons--cm$^{-2}$--s$^{-1}$--keV$^{-1}$
to better reveal the low energy turnover evolving to a slight excess;
and for detailed comparisons with other published X--ray spectra.
Following Strohmayer et al.\markcite{st1} 1998,
in Figure 4.g we plot the best--fit model spectra
but normalized to unity at 10 keV,
to better illustrate the evolution of just the spectral shape.
Notice the fraction of low energy emission consistently rises
as the burst progresses.
We list the parameters
from empirical fits to these data in Table 1.
At no time was there evidence for a line feature, either in emission or absorption.
For all the spectra, adding a line described by three parameters
(width, intensity, and line energy) reduced the value of
$\chi^2$ by less than 3;
and adding a line described by two parameters
(width and intensity, with the energy fixed at that of one of the {\it Ginga}/LANL
lines;
Murakami et al.\markcite{mu1} 1988;  Fenimore et al.\markcite{fe1} 1988) 
lowered $\chi^2$ by less than 2.

\placefigure{figure4}

The first spectrum contains the rise and maximum of the first peak.
Using just an absorbed power--law model, one finds it is very flat
(photon index 1.10$\pm$0.4 from 2--50~keV), and
featureless out to a slight apparent steepening of the
spectrum that starts at around 45~keV.  This is too flat
to accommodate a thermal synchrotron model, or
even a sum of hardening thin thermal
bremsstrahlung spectra,
or of black body spectra.
The turnover at low energies is most pronounced,
consistent with a high average absorption of
0.3$\pm$0.05$\times$10$^{23}$ cm$^{-2}$
(estimated usng the power--law model with no black--body component).
This is greater,
by over an order of magnitude,
than the total interstellar column density
to the edge of the Galaxy in this direction
(Lucke\markcite{lu1} 1978;  Kerr et al.\markcite{ke1} 1986;  
Lang\markcite{ln1} 1980).
This low energy turnover is not present in any of the spectra
from later time intervals, including the second peak.
All show column densities consistent with zero, and in fact
show evidence for a slight excess above the power--law.

If one scrutinizes the $<$5~keV
spectra from other {\gray} bursts, one notices a similar
dearth of low energy photons at the time of the first peak,
although this is not always present
(Strohmayer et al.\markcite{st1} 1998; Murakami et al.\markcite{mu3} 1992
and references therein; 
Katoh et al.\markcite{ka2} 1984).
We note this is consistent with ``evolving low energy cutoffs''
observed at higher energies on the rising edge of some peaks
of other {\gray} bursts 
(Strohmayer et al.\markcite{st1} 1998; Kargatis et al.\markcite{ka} 1994; 
and Pendleton et al.\markcite{pe1} 1996).
However it is interesting to note that when this low-energy turnover exists
in a {\gray} burst,
it apparently does not always occur on the rising edge of the first peak.
There is a dramatic example in Katoh et al.\markcite{ka2} (1984) 
of extreme low energy
suppression in the middle of a burst.

\placetable{table1}

As each of the two main
peaks decays, the high energy tail steepens,
while the peak of the
emission shifts to lower and lower energies.
This ``black body'' portion of the low energy emission comprises
$\sim$10\% of the $<$60~keV emission and
$\sim$2\% of the total emission.
For GRB~780506, we estimate the 3--10~keV emission
to be $\sim$6\% of the energy emitted $>$30~keV,
during just the first peak of the burst;
when we included both peaks of the burst,
we found roughly 7--20\% of the total burst energy was
emitted between 3--10~keV,
with the uncertainty due chiefly to
the uncertainty in the shape of the spectrum $>$50~keV
as the burst progressed.
This is higher than similar ratios
calculated for 
most of the other {\gray} bursts for which it has been
measured
($\sim$1\%--3\%,
from Laros et al.\markcite{la1}\markcite{la2} 1984a,b;
$\sim$0.5\%, with an uncertainty of a factor of 5,
from Figure 6 of Katoh et al.\markcite{ka2} 1984; and
$\sim$1\%--10\% from Table~1 in 
Strohmayer et al.\markcite{st1} 1998,
for the four bursts with known positions).
From comparisons of the spectral shapes,
one finds this is due to less $>45$~keV emission in GRB~780506
(i.e. steeper than average power--law at higher energies)
rather than softer or less highly absorbed low energy emission.

Although Murakami et al.\markcite{mu1} (1988) preferred to interpret
the softening spectrum in the decay portion of each peak
as cooling black-body emission,
in GRB~780506,
a power law tail was visible,
ruling out both simple superpositions
of thermal bremsstrahlung and black body spectra,
for as long as there were sufficient
counts per energy bin to distinguish among spectral models.
This is in fact consistent the {\it Ginga} X--ray observations 
(Murakami et al.\markcite{mu1} 1988;
 Yoshida et al.\markcite{yo1} 1989;
Murakami et al.\markcite{mu3} 1992).
Earlier measurements of X--ray emission did not have good spectral data
below several keV, so comparisons were not feasible.

Imamura and Epstein\markcite{ni1} (1986) and Epstein\markcite{ep1} (1986)
pointed out that,
even under the assumption that most of the X--ray emission
comes from thermal re-processing
(for example absorption and re-radiation by nearby matter),
only a small fraction ($<$2\%)
of the total {\gray} energy could
have been `thermalized',
unlike what one expects from an explosive event taking place
close to the surface of a neutron star
(the well--known `X--ray paucity' constraint).
We point out that, at least for this event, the spectra preclude
a simple thermal origin for the X--ray emission.
Although the ratio of 3--10 keV emission to $>$30 keV emission
was quite high ($\sim$7--20\%) for this event,
the fraction that could have come from a black-body component
is constrained to be 2\% or less.
We note the total X---ray emission from GRB~970228 was similarly high
(Costa et al\markcite{co97} 1997).

It is difficult to compare the $<$few keV behavior of GRB~780506
with that of many {\it Ginga} bursts, as the angle of the burst to the
detectors was unknown for many events, and
uncertainty in response is greatest at these lowest energies.
A few bursts were localized by other spacecraft.
Of these, in GRB~900126 (localized by WATCH), a low energy turnover,
consistent with absorption, is visible;
while in GRB~910429 (localized by BATSE) there is at best a suggestion
of flattening below a few keV.
In BATSE data, sometimes the lowest channel of the spectroscopy detectors
measured as low as a few keV.
Out of $\sim$100 bursts, Preece et al.\markcite{pr1} (1996) found
$\sim$10--15\% showed low energy emission significantly in excess above
the spectrum extrapolated from higher energies;  and about an equal number
with low energy roll--offs.

The strong low energy turnover seen
at the onset of GRB~780506
serves as a reminder that soft X--ray photons might be removed
from an observed burst spectrum
by processes where the cross section is much higher
for soft X--ray photons than for {\grays}, such as
photo-electric absorption;
Compton scattering in extended cold material;  or
magnetic field processes such as ``cyclotron scattering".
Liang\markcite{li1} (1994) has proposed a cold absorption model
where the absorbing material is Fe--rich.  Although the GRB~780506 spectra
are consistent with this simple picture of hard--to--soft evolution 
 due to an optically thick absorbing layer being heated or blown away,
there is no evidence of the predicted absorption edge at $\sim$7~keV
or K-alpha emission feature at $\sim$6.7~keV.  
Brainerd\markcite{bra94} (1994) has proposed a Compton attenuation model
which both suppresses the low energy emission and predicts an 
X--ray afterglow.  The {\gray} burst is assumed to be attenuated
by scattering and absorption through dense molecular clouds.
However the absorption edge would be red--shifted; and
with this model, there is not a natural explanation for how
the heaviest absorption should be on the rising edge of a burst.
McBreen, Plunkett, and Metcalfe\markcite{mcbplumet93} (1993)
invoke a thick cocoon of material around a source of a relativistic
beam of radiation; and show a slow--down of the beamed material
producing an evolving beaming factor, to get the characteristic
hard--to--soft spectral evolution.  But the optical depth
does not change in their scenario,
so it is difficult to see how an evolving low--energy turnover 
would be produced.
Liang et al.\markcite{li97} (1997) explicitly model this
characteristic spectral evolution with an expanding, cooling
plasma for which the Thomson depth decreases as the plasma thins.
However the $<100$~keV portions of their spectra do not seem
to match the flattening that was seen here.

The dramatic recent detections of several optical afterglows,
one possibly associated with a galaxy 
(GRB~970228; van~Paradijs et al.\markcite{van} 1997) 
and one with absorption lines (GRB~970508; Metzger et al.\markcite{metzger} 1997) 
apparently indicating a redshift
of $Z\sim0.83$, 
have focused attention on cosmological
expanding fireball models (M\'esz\'aros and Rees\markcite{me97} 1997,
henceforeth MR97; Sari and Piran\markcite{saripiran} 1996; and references therein).  
These are modeled as relativistic
shocks, with the burst being the signature of shocks developed internal to
 the ejecta wind (e.g., Pilla and Loeb\markcite{pillaloeb} 1998); 
and longer time--scale
emission from external shocks produced as the ejecta plows
into the local interstellar medium.
This is a very appealing picture, but again we note that in detail,
our $<50$~keV spectra do not match at least these preliminary predictions.
In these models,
for the canonical GRB spectrum we described in the third paragraph
of this section,
the break energy at hundreds of keV is modeled as either due to
inverse Compton (IC) or to the synchrotron peak (MR97; Katz et al 1997).
Under very general assumptions, for a relativistic shock, the photon
power--law index for the spectrum below the synchrotron peak
should asymptotically approach 0.7 -- 1.5 .  Cohen et al. (1997),
using the ($>20$~keV) spectra of 11 bright BATSE bursts, assume the
few hundred keV break to be the synchrotron peak, and find moderate
consistency.  However, following their method of fitting the lowest
energy channels by eye (but we used photon power--law spectra, while
they used $F_{\nu}$ spectra), 
we find, for the average spectrum
of each peak (Figure 3), asymptotic slopes of $\sim$ 0.5 and 1.9
(equivalent to $-1.5$ and $-0.1$ in $\nu F_{\nu}$ units).
For the individual 10.24~s spectra (Figure 4) we find asymptotic
slopes of $\sim ~ -0.7$, 0.5, 0.3, 0.7, 2.3, and 3.3; which
are widely outside these limits.
One could argue that the initial low energy turnover is due to an evolving
synchrotron peak (MR97), as it is emphasized by all authors
that its position is unknown (Cohen et al 1997, Katz et al 1997).
In this case the higher energy peak would be due to inverse Compton
(which is also what Brainerd\markcite{bra94} 1994 suggests).
However we note that among the ensemble 
of historical measurements of X--rays
from {\gray} bursts
there are many with shapes outside the predicted
low--energy limits
(Laros and Nishimura\markcite{la3} 1986); and
some, such as GRB~811016, have strong low energy suppression
in the middle of the burst, precluding a simple evolving synchrotron 
interpretation
(Katoh et al.\markcite{ka2} 1984).
There are now more complex kinds of cosmological theories:
for example, Katz\markcite{Katz97} (1997)
emphasizes the difficulties of realistically producing
{\grays} with complex time--histories in these scenarios,
and instead suggests pulsar--like emission.

From just these {\it HEAO~1} A--2 data, it is hard to distinguish among:
$N_H$ absorption plus Compton attenuation, followed by black body emission
or soft X--rays forward--scattered by dust
(Brainerd\markcite{bra94} 1994);
$N_H$ absorption plus Compton down--scattered flux 
(Yaqoob\markcite{yaq96} 1997); 
a moving synchrotron peak from a relativistic expanding fireball model; 
as well as the effect of a possible red--shift on the absorption edge.
Future experiments with good energy resolution below a few keV
may be able to distinguish among these (Forrest et al.\markcite{fo2} 1995).

\subsection{ Afterglow Light Curve }

In Figure 5 we display the light--curve of just the X--ray afterglow,
in 122.88~s bins, along with the
A--2 HED3 Layer 1 2--20~keV rates for the same times.
One sees a peak at about 7 minutes
($\pm 1$ minute) after burst onset.  When exponential models
are used to fit the rise (first two bins plus the peak)
and decay (peak plus the last thirteen points in this orbit),
one finds time--scales of $190\pm80$~s with $\chi^2 = 0.3$ for 1 d.o.f.,
and $2090\pm 1430$~s with $\chi^2=21.$ for 12 d.o.f.,
respectively.  We note that the `decay' portion was not well--fit
by this simple model.
We suggest that afterglow variability
on time--scales on the order of $\sim {{1}\over{2}}$ hr is potentially
interesting, and should be investigated in the future.
We also tried plotting the burst and afterglow on a log--log
scale to see if there was a discernible power--law decay envelope
(Costa et al. 1997), but this only seemed to accentuate
the variability of this event.
In particular,
Costa et  al.\markcite{co97} (1997) extrapolate their GRB~970228 afterglow
measurement to be part of a power--law decay
of the burst itself.  In contrast, here, there seemed to be a clear break
between the X--rays from the burst and the rise to the afterglow:
a different phenomenon than
the decay of an X--ray tail
lasting minutes observed by Vela (Terrell et al.\markcite{te2} 1982)
and {\it Ginga} (Yoshida et al.\markcite{yo1} 1989;
Murakami et al.\markcite{mu3} 1992).

At shorter timescales,
an FFT of this time interval, using 5.12~s time bins,
yielded a 90\% upper limit
of $<$ 2\% on any pulsed component less than 15 minutes.

\placefigure{figure5}

Qualitatively, one might argue that this afterglow could be
the signal of the relativistic plasma that produced the burst
encountering the local interstellar medium
(MR97).
Alternatively, Katz et al.\markcite{kapisa97} (1997) considers
the source sputters (in gamma-rays as well as lower energies)
at a much lower level for on the order
of a day before and after the bright ``burst" portion.
Quantitatively, however, these models still have many unknowns.
We note the afterglow spectrum is too soft to be
an extrapolation of flux below the synchrotron peak,
for example (Cohen et al.\markcite{cohen97} 1997 and references therein).
Future high signal--to--noise measurements of spectra
and light--curves during this interesting time,
from directly after a burst until several hours after,
would be useful in shaping these models
(Smith et al.\markcite{smith97} 1998; 
Takeshima et al.\markcite{ta97} 1998).

We asked the question, what were our limits on longer timescale
(several hour) emission
from the burst source,
excluding the afterglow?
When times of high particle background were excluded
(MacIlwain's $L$--parameter restricted to $\le 1.4$),
we obtain a 2$\sigma$ upper limit
on persistent emission from the burster
of $<$ 3 millicrabs (0.005 cts-cm$^{-2}$-s$^{-1}$)
using scanning data for several days after the burst and afterglow.
We found no evidence of significant emission before the burst,
but point out that the Earth's disk would have
occulted any emission from the source between
$\sim {{1}\over{2}}$ hr to 10~s before the beginning of GRB~780506.

At higher energies ($\ge$100~keV),
Klebesadel et al.\markcite{kl1} (1984),
Klebesadel\markcite{kl2} (1992), and
R. Klebesadel (1985, private communication),
describe an event observed by PVO, GRB 840304, composed of 2 intense,
$\sim$30~s FWHM peaks, 100~s apart, which were followed
by $\sim$1500 s of emission
 about two orders of magnitude fainter than the peak intensities.
This extended emission had roughly an $E^{-1.3}$ spectrum,
much harder than the emission here.
Klebesadel (1989) points out that the $\sim$1500~s duration is only
a $\sim3\sigma$ deviation from the long duration tail of
the distribution of event durations observed by PVO.
At even higher energies, Hurley et al.\markcite{hu1} (1994) describe
extended GeV emission following GRB~940217, and remark on
GeV emission detected $\sim$minutes after four other events.
Since the spectra of these reports of extended emission
are not consistent, it is not clear whether they are related phenomena.

\section{ Conclusion}

In the past, longer time--scale X--ray emission from {\gray}
bursts has been seen as an indicator of a separate thermal component
(Laros and Nishimura\markcite{la3} 1986;
Chernenko and Mitrofanov\markcite{ch1} 1994).
We argue above that the increase in X--ray time--scales follows the
same pattern observed at {\gray} energies, and
that the energy in 2--60 keV X--rays
is released through the same non-thermal processes that generate 
the {\gray}
emission.
This pattern of spectral variability is visible in the X--ray spectra
as well as the light--curves.
We have argued that when viewed in detail, the 2-60 keV X--ray emission
shows no unambiguous evidence for a thermal interpretation.
We set a limit of $<$2\% on the black--body component of the emission.
Unlike some {\it Ginga} {\gray} bursts, there was no evidence for
line--features.
Instead, the general trend seems to be a non--thermal 
power--law in the X--rays
below the characteristic break energy or maximum in $\nu F_{\nu}$
(typically $\sim$200~keV but with a wide range; here, $\sim$45~keV).
Below a few to $\sim$10~keV this evolves to either a turnover consistent
with high intrinsic absorption or excess soft emission
(Strohmayer et al.\markcite{st1} 1998; Preece et al.\markcite{pr1} 1996).
There was no evidence to suggest the second peak was
part of an afterglow that was distinct from the burst itself,
unlike GRB~970228 (Costa et al.\markcite{co97a} 1997; 1998).
We note that, like many bursts measured previously in X--rays,
the spectral shape varies widely outside low--frequency limits
predicted by current relativistic shock models.

Two characteristics of GRB~780506 may be unusual.
One is the softer than average $>$45~keV spectrum, placing
it in the ``NHE'' subclass
(see Pizzichini et al.\markcite{pi95} 1995; 
Kouveliotou et al.\markcite{koetal96} 1996;
Belli\markcite{be97} 1997; and Pendleton et al.\markcite{pe97} 1997).
The more dramatic is its faint, $\sim$1~keV afterglow.
Extended emission has now been detected from at least seven of
nine bright {\gray} bursts that have had extensive
multi--spacecraft follow--up 
(Paczy\'nsky and Kouveliotou\markcite{pa97} 1997 and references therein;
Heise et al.\markcite{heise97b} 1997; 
Piro et al.\markcite{co97b}\markcite{98}1998a,b;
Halpern et al.\markcite{halpern} 1997).
However these detected apparently fading emission,
on timescales of hours or days,
not the initial resurgence of soft X--rays reported here.
Also, though extended emission had been detected at higher energies
($>$100~keV, 
Klebesadel et al.\markcite{kl1} 1984; Klebesadel\markcite{kl2} 1992;
$>$100~MeV, Hurley et al.\markcite{hu2} 1994), it is not clear these are
the same phenomena.

For the afterglow,
it is not clear whether it 
is the rare, long--duration tail of the overall {\gray}
burst duration distribution (Klebesadel\markcite{kl2} 1992; 
Kouveliotou et al.\markcite{ko1} 1994);
or whether the $\sim$1~keV emission, few minute rise time,
irregular flux and potential ${{1}\over{2}}$ hr decay
are characteristic of most 
{\gray} bursts.
This is a new regime.
First, earlier instruments with wide fields of view also tended to 
have high backgrounds
due to source confusion (see Laros and Nishimura\markcite{la3} 1986
and references therein).
Second, burst monitors such as those on {\it BeppoSAX} have historically
stored data for only a limited time after a burst
(e.g. Boella et al.\markcite{boella} 1997).
Recently, the {\it RXTE} ASM and the WFC on board {\it BeppoSAX}
have begun to accumulate data on
 {\gray} bursts
observed at least minutes after burst onset,
and may be able to set constraints in the near future
(Smith et al.\markcite{smith97} 1998; 
Takeshima et al.\markcite{takeshima97} 1998;
Piro et al.\markcite{piro98} 1998a,b).
If this resurgent X--ray afterglow 
is a feature common to many {\gray} bursts,
and if it is accompanied by optical emission,
then there may be a $\sim$60 minute window (rather than a one minute window)
during which it is possible to observe
brighter optical emission associated with a {\gray} burst.

For the burst itself,
it remains for future experiments with good signal--to--noise 
below a few keV such as {\it HETE}
(Ricker et al.\markcite{ri1} 1992) and
{\it CATSAT} (Forrest et al.\markcite{fo2} 1995);
or more extended observations of GRBs from current
satellites such as {\it RXTE} (Takeshima et al.\markcite{ta97} 1998;
Smith et al.\markcite{smith97} 1998)
and the Italian--Dutch satellite {\it BeppoSAX}
(Boella et al.\markcite{boella} 1997)
 to determine
to what extent the generally non--thermal spectrum;  
the hard--to--soft variability;
and the low energy turnover evolving to slight excess 
$\sim$1 keV emission;
are characteristics of X--rays from all {\gray} bursts.

{\acknowledgments
  The GRO/COMPTEL group at UNH's
Institute for the Study of Earth, Oceans and Space, under J. Ryan,
provided support throughout.
This work was supported through NASA grant NAG~5-1753, which included funds
for the purchase of a workstation.
This research has made use of data obtained through the 
High Energy Astrophysics Science Archive Research Center Online Service,
provided by NASA--Goddard Space Flight Center.
}

{
\appendix
\section{Two Bayesian Methods for Finding the Position of a Variable Source}

\subsection{Overview}

We have derived and applied two Bayesian likelihood ratios for
the problem of finding the position of a time--varying source from
comparisons of detector rates.  This Bayesian formalism allowed one to
integrate over the unknown light--curve shape, parametrized by $\{s_k\}$,
freeing one to
handle datasets with many more time bins, in which the true source
intensity and positions of the detector axes were allowed to vary.  
Loredo\markcite{lo1}\markcite{lo2} (1990, 1992) and 
Gregory \& Loredo\markcite{gl1} (1992; henceforth GL92)
suggest that Bayesian methods can extract
the maximum amount of information 
when carefully applied to data analysis problems.
The first Bayesian likelihood ratio,
for a high signal--to--noise Gaussian approximation,
was simpler and more efficient.  
In the Poisson case, necessary for faint sources,
although the likelihood ratio was more complex and the calculation
more computer--intensive, 
the constraints on position were tighter.
It should be straightforward to apply this same process,
to data from other detectors,
where Bayesian methods could be expected to give an improved result.

Sampling statistics deals with data--space, with $p(D \vert{\bf \Omega}, \hyp, I)$,
the {\em direct probability} of the data $D$ 
given an hypothesis or model $\hyp$
with parameters ${\bf \Omega}$ and prior information $I$.
In Bayesian inference one works in parameter--space, with the probability
of the parameters ${\bf \Omega}$ given the model $\hyp$ 
plus prior information $I$,
$p({\bf \Omega} \given D, \hyp, I)$.
Formally, one calculates one from the other via Bayes's theorem,
(a consequence of the product rule of probability):
\begin{equation}
\label{bayes-thm}
p( {\bf \Omega} \given D, \hyp, I) ~=~ 
p({\bf \Omega} \given \hyp, I) p( D \given {\bf \Omega}, \hyp, I) 
  / p( D \given \hyp,  I )
\end{equation}
The concept is that one {\em updates} the {\em prior probability}
$p({\bf \Omega} \given \hyp, I)$ with the {\em direct probability}
of the data $p( D \given {\bf \Omega}, \hyp, I)$, normalized with
$p( D \given \hyp,  I )$, to get the {\em posterior probability}
$p( {\bf \Omega} \given D, \hyp, I)$.

One benefit of working in parameter space is the ability to integrate
over uninteresting parameters.
For the application presented here, one integrates
analytically over unknown light--curve parameters,
and numerically over the measured background rates,
to obtain a likelihood ratio parametrized only by the unknown source
position $(\phi, \theta)$.
The final output is a map of this Bayesian likelihood ratio as a
function of $(\phi,\theta)$.  
Position constraints are found by drawing contours of constant
log probability around regions of highest probability, or credible regions
(GL92; Loredo 1990; Loredo 1992).

The two algorithms we derived, high and low signal--to--noise 
likelihood ratios,
were tested on {\it HEAO~1} A--2 proportional counter data from
 sources with known positions.
However, the algorithms should be applicable to any source observed by
multiple detectors with overlapping, well--understood spatial responses; similar spectral responses; 
and predictable backgrounds.  
These algorithms 
were then applied to both GRB~780506 and the
$\sim$50 minute 2--20 keV emission observed following it.
Originally, Connors\markcite{co1} (1988) had tried a maximum likelihood 
method.
The improved position constraints from the Poisson Bayesian likelihood ratio
strengthen the hypothesis that this emission
is associated with the burster itself, rather than coming from a 
serendipitous source (Connors and McConnel\markcite{co96} 1996).

\subsection{ Data}

The position--finding algorithms derived here were first tested on
a variety of {\it HEAO~1} A--2 data: scanning and pointed data; 
on bright and faint sources.  
These are listed in Table~2.
During the Crab observation, 
the detector look--direction rotated at about 1\arcdeg ~
every 5~seconds.
During the other observations the detector axes rocked by about
1\arcdeg ~ on timescales of minutes.
Since the derivation assumes the
background to be constant, we chose data that was the least sensitive
to particle contamination: MED Layer~1 rates, 
restricted to times of high rigidity 
(MacIlwain's $L$-parameter restricted to $L<1.2$).

These algorithms were then applied to sources with unknown positions:
GRB~780506 and the following soft X--ray emission.

\placetable{table2}

\subsection{Previous Method:  Maximum Likelihood }

The earlier position--finding algorithm 
(of Connors\markcite{co1} 1988) was
a maximum likelihood method,
with the position $(\phi,\theta)$ of the source; 
the rate $\{s_k\}$ in each time bin;
and the (constant) background $b_j$ in each field of view, as parameters.
Let $\epsilon_{jk}(\phi,\theta)$
represent the detector effective area of the $j^{th}$ detector field of view
as a function of source position $(\phi,\theta)$ at time $t_k$.  
Then the total expected rate $\mu_j$
in the $j^{th}$ detector field of view at time $t_k$ can be modeled as:
\begin{equation}
\label{mujk}
\mu_{jk}(\phi,\theta) ~=~  b_j + s_k \epsilon_{jk}(\phi,\theta) .
\end{equation}

In Connors (1988), the background parameters $b_1$ and $b_2$
were determined from previous measurements; and the detector effective areas
$\epsilon_{jk}$ were determined from calibrations with known sources;
so only the source position $\phi,\theta$ and its intensity as a function
of time, as parametrized by the $s_k$, were unknown.  Following the
maximum likelihood approach of Cash\markcite{ca1} (1978), 
the light--curve parameters $\{ s_k \}$ are allowed to 
freely float until
one finds the maximum probability for that $\phi,\theta$ position.  
However the maximum likelihood method required 
independent fit--parameters for the source intensity in each time bin.
For transients with durations of order roughly minutes, this meant at worst
dozens of light--curve parameters.  For sources observed for
an hour, one would need closer to $10^3$ independent parameters.   
This was not practical. 
One could bin the data into longer time bins;
however since even during pointed observations, the detector point--direction
varied, positional information could be lost.

This was the case for the $\sim$50 minute emission that followed
GRB~780506.

\subsection{ First Bayesian likelihood ratio:  high signal--to--noise }

In the previous section, both the background parameters $b_1, b_2$ and
the unknown light--curve parameters $s_k$ could have
been considered
``nuisance'' parameters that made finding constraints on the source position
computationally difficult.  In this section, one uses a Bayesian approach to
integrate over them, with the simplifying assumption 
that a Gauss--Normal distribution is a reasonable approximation
of the Poisson distribution.
This works well when the signal is at least several $\sigma$ above
the background in each time bin.

First, one specifies prior probabilities for each
of the parameters, $p( s_k \given I), p(b_1 \given I),$ and $p(b_2 \given I)$.
When there is no prior measurement, for parameters describing a rate,
GL92 uses an invariance argument to suggest a uniform
prior over a range $s_0$ to $s_1$, where 
$s_0$ might be zero, and $s_1$ the maximum rate the detectors can 
measure:
$
p(s_k \given \hyp, I ) ~=~ { {1} \over {s_1} }.
$
For the backgrounds $b_1$ and $b_2$, there are prior measurements:
$B_1$ and $B_2$ counts in times $T_1$ and $T_2$,
in detector fields--of--view 1 and 2, respectively.  Therefore the appropriate
expressions for the prior probabilities on $b_1$ and $b_2$ are
(Loredo\markcite{lo1} 1990)
\begin{equation}
\label{b12-priors}
p( b_1 \given B_1, T_1, I ) ~=~
 {\euler}^{-b_1 T_1} { { (b_1 T_1)^{B_1} } \over {B_1 !} }
~{\rm ~~ and ~~ }~
p( b_2 \given B_2, T_2, I ) ~=~
 {\euler}^{-b_2 T_2} { { (b_2 T_2)^{B_2} } \over {B_2 !} }.
\end{equation}
One combines them with the direct probability, using Bayes's Theorem,
to obtain the posterior probability. 
For this section, the direct probability is assumed to be Gaussian.

First, one deals with the $s_k$.
One integrates (marginalizes) analytically over each $s_k$, 
then takes the product
over all $k$ data bins, with $y_{jk}$ counts in each data bin,
the mean given by equation A2,
and its associated $\sigma_{jk}$ given by $\sigma_{jk} = \sqrt{y_{jk}}$.  
In the limit
where the Gaussian dsitribution in $s_k$ is sharply peaked compared to the
width of the integration interval $[0,s_1]$, one obtains a simplified
expression for the (negative) log of the posterior probability
$\bodds( b_1, b_2, D1, D2 \given ~ \phi, \theta, \hyp ) \propto
-\log \bigl[ p( b_1, b_2, D1, D2 \given \phi, \theta, I ) \bigr]$:
\begin{equation}
\label{bodds-def}
\bodds(  b_1, b_2, D1, D2 \given ~ \phi, \theta ) ~=~
$$
$$
- \Bigl(
{ B_1 - b_1 T_1} + {B_1} \log\bigl( { {b_1 T_1} \over {B_1} }\bigr)
 \Bigr) -
\Bigl(
{ B_2 - b_2 T_2} + {B_2} \log\bigl( { {b_2 T_2} \over {B_2} }\bigr)
 \Bigr)
 + {1\over2} \sum_{k=1}^{N}
 { {\bigl( \Delta_{1k}  -  \Delta_{2k} \bigr) ^2 } 
    \over { {\sigma_{Tk}}^2 }
  },
\end{equation}
where
\begin{equation}
\label{sigTdef}
\sigma_{Tk} ~\equiv~ 
       \sqrt{
              \bigl(\sigma_{1k}\epsilon_{2k}\bigr)^2 +
              \bigl(\sigma_{2k}\epsilon_{1k}\bigr)^2
             },
\end{equation}
and
\begin{equation}
\label{Del12def}
\Delta_{1k} ~\equiv~ \bigl( y_{1k} - b_1 \bigr) ~ \epsilon_{2k};
 ~~~ {\rm and} ~~~
\Delta_{2k} ~\equiv~ \bigl( y_{2k} - b_2 \bigr) ~ \epsilon_{1k}.
\end{equation}

If one or more of the detector effective areas $\epsilon_{jk}$ were zero,
the appropriate term(s) in the sum 
would be replaced by:
\begin{equation}
\label{log-term-1zero-gauss}
 \Bigr({{y_{2k} - b_2}\over{\sigma_{2k}}}\Bigl)^2 ~~{\rm for~}\epsilon_{1k}=0;
 {\rm ~~or~~}
 \Bigr({{y_{1k} - b_1}\over{\sigma_{1k}}}\Bigl)^2 ~~{\rm for~}\epsilon_{2k}=0;
\end{equation}
or by
\begin{equation}
\label{log-term-2zero-gauss}
\log(s_1) + \Bigr({ {y_{1k} - b_1}\over{\sigma_{1k}} }\Bigl)^2 +
            \Bigr({ {y_{2k} - b_2}\over{\sigma_{2k}} }\Bigl)^2 
\end{equation}
for $\epsilon_{1k} = 0$ and $\epsilon_{2k} = 0$.

Second, one numerically integrates the negative exponential of this form
 over $b_1$ and $b_2$ to obtain
$\like( D1, D2 \given ~ \phi, \theta, \hyp ) \propto
\log \bigl[ p( D1, D2 \given \phi, \theta, I ) \bigr]$,
the log--likelihood of the data given source positions $(\phi,\theta)$.
For each position in a grid of $(\phi, \theta)$,
one first found the minimum of $\bodds$, $\bodds_0$; and then
numerically integrated $\exp\{ -\bodds + \bodds_0 \}$ over the background
parameters $b_1$ and $b_2$.
The minimization used a modified Levenberg--Marquardt algorithm,
which returns both a best--fit for each parameter and an estimate of its 
$\sigma$, using the curvature to find a Gaussian approximation.
(Bevington\markcite{be1} 1969; 
Press {\it et. al.}\markcite{pr1} 1986 and references therein).
The numerical integration used Gaussian quadrature, with Gaussian weights,
assuming the $\sigma$ found in the previous step,
and using Hermite polynomials 
(Scheid\markcite{sc1} 1968;  Abramowitz \& Stegun\markcite{ab1} 1972;
Press {\it et. al.}\markcite{pr1} 1986), typically with eight roots.

For the data tried here, the whole procedure was relatively fast, 
requiring about two hours of Decstation~5000/125,
(or one hour of 40~MHz Sparc~10) CPU time for 852 time bins
and $10^3$ grid--points.
The computer time scaled linearly with the number of 
time bins in the data--set and grid--points requested.

Third, one 
maps out contours of constant 
$p( \phi, \theta \given  D1, D2, I )$, assuming a uniform prior
on $\phi, \theta$,
to obtain constraints on the parameters.
Position constraints are delineated by contours of constant log probability
drawn to contain regions of highest posterior probability,
or Credible Regions (Loredo\markcite{lo92} 1992 and references therein).
We used Credible Regions containing 95.45\% and 97.23\% of the posterior probability to delineate the source positions.

\subsection{Application to Data: Known and Unkown Sources }

This high signal--to--noise algorithm was tested on data from several
known point sources.
The first panel of Figure 6 shows the results, 
plotted in two grayscale contours,
for 5 scans (182 time bins) 
across the Crab nebula plus pulsar (on D.O.Y.~1977 260).
The second panel shows the results from 2.5 hr (632 time bins) from
the pointed observation
of the highly variable neutron--star binary GX~301-02 (on D.O.Y.~1978 25).
The last two panels show the results on a pointed observation of
the cluster Abell~401 (D.O.Y.~1978 39).
The left panel displays the results on the entire point (832 time bins),
while the right shows the results for a single orbit (352 time bins).
The latter is analogous to the data 
from the unknown source following GRB~780506,
for which data from a single orbit was used.
For each plot, the darker contour encloses a 99.45\% (2 $\sigma$)
credible region, and the lighter a 99.73\% (3 $\sigma$) credible region.
The true source positions are marked with an asterisk.
It is clear that the high signal--to--noise algorithm works for
the bright test cases, but is a little off (although still within 3 $\sigma$)
for the faintest source, Abell~401.
This will be addressed when using the full Poisson method, 
for lower signal--to--noise data.

\placefigure{figure6}

Next, this algorithm was used to constrain the position of GRB~780506 and
the position of the unknown transient that occurred just afterwards.
These are {\it HEAO~1} A--2
5.12~s rates from the first layer of the offset High Energy Detector,
HED1, with data constrained to MacIlwain's $L$--parameter $L < 1.4$,
obtained during the first orbit following GRB~780506
(Figures 1 and 3).
For Figure 7, the GRB~780506 95.45 and 99.73\% credible regions
were plotted in gray contours, 
while those for the extended emission were overplotted
as line contours.
The error box for GRB~780506 changed little from that in 
Connors\markcite{co1} (1988).  
However, the position
constraints for the fainter transient changed appreciably.
The Bayesian algorithm allowed for variations in intensity in each time bin,
while because of the large number of intensity parameters that would have
been required, the likelihood method could not.  This suggests 
the position contours displayed in Figure 7 are the more physically 
realistic constraints.  However, recall that in two places in the derivation,
one assumed the signal was at least several $\sigma$ above the background
in each time bin.  This is not strictly true for the faint $\sim$50 minute
emission.  In the next section, we sketch out the analogous derivation,
but without that simplifying assumption.

\placefigure{figure7}

\subsection{ Second Bayesian likelihood ratio:  low signal to noise }

Here, one follows the same steps as in the previous section:
writing down the joint sampling probability; using the same priors,
marginalizing analytically over the ${s_k}$;  then numerically over the
background rates;  but for the full Poisson expression necessary
when measuring faint sources.
In order to do this, one defines a numerically convenient
log--likelihood analogous to equation (\ref{bodds-def}).  The output will
be the log of a global likelihood ratio.

These expressions will be more difficult to compute than for the Gaussian
approximation above, but 
provide better constraints for the low signal--to--noise cases.

Instead of a Gauss--Normal approximation for the sampling (direct) 
probability,
one starts with the Poisson expression for the joint probability 
for a single time bin.
One again uses a uniform prior on the $s_k$, 
marginalizes over each $s_k$, and takes the product over all $k$ time bins.
The log of the resulting expression
likelihood is not as simple as that for the Gaussian approximation,
but straightforward to calculate:
\begin{equation}
\label{likli-poiss}
\bodds (D1, D2 \given b_1, b_2, \phi,\theta, I) ~=~
\Bigl(b_1 T_1 - B_1 + B_1 \log\bigl[ {{B_1}\over{b_1 T_1}} \bigr] \Bigr) +
\Bigl(b_2 T_2 - B_2 + B_2 \log\bigl[ {{B_2}\over{b_2 T_2}} \bigr] \Bigr) +
N ( b_1+b_2 ) 
$$
$$
 - \sum_{k=1}^{N} \log\Biggl(
\sum_{l_{1k}=0}^{y_{1k}}  \sum_{l_{2k}=0}^{y_{2k}}
{ {y_{1k}} \choose {l_{1k}} } { {y_{2k}} \choose {l_{2k}} }
{b_1}^{y_{1k}-l_{1k}}  {b_2}^{y_{2k}-l_{2k}}
{ 
  { {\epsilon_{1k}}^{l_{1k}} {\epsilon_{2k}}^{l_{2k}} }
      \over
    { (\epsilon_{1k} + \epsilon_{2k})^{l_{1k}+l_{2k}+1} } } ~ (l_{1k}+l_{2k})!
  \Biggr),
\end{equation}
for $\epsilon_{jk}$ non--zero; with the appropriate term in
the sum over $k$ replaced by
\begin{equation}
\label{log-lambda-1zero-poiss}
   y_{2k} \log(b_2) +  \log\Biggl(
\sum_{l_{1k}=0}^{y_{1k}} { {y_{1k}} \choose {l_{1k}} } {b_1}^{y_{1k}-l_{1k}} 
{ {l_{1k}!} \over {\epsilon_{1k}} } \Biggr),
\end{equation}
if $\epsilon_{2k} = 0$ (or its equivalent if $\epsilon_{1k} = 0$);  or
\begin{equation}
\label{log-lambda-2zero-poiss}
   y_{1k} \log(b_1) +    y_{2k} \log(b_2),
\end{equation}
if both $\epsilon_{1k} = 0$ and $\epsilon_{2k} = 0$.

Again, for each $\phi,\theta$ bin, 
$\min\Bigl(\bodds(D1, D2 \given b_1, b_2, \phi,\theta, I)\Bigr) = \bodds_{0}$ 
was found, and the exponential of 
$(-\bodds(D1, D2 \given b_1, b_2, \phi,\theta, I) +\bodds_{0})$
was numerically integrated over $b_1$ and $b_2$.

Unlike for the Gaussian approximation, these calculations were very slow,
requiring about $\sim$70 times as much CPU time.
The computer time scaled linearly not only with the number of 
time bins in the data--set and grid--points requested,
but with the average count--rate as well.
In practice, then, one prefers to use this algorithm for
short datasets with low count--rates.
However, for these low signal--to--noise observations, the restriction
of $\{s_k\} \ge 0$ that is implicit in the Poisson formulation
led to tighter and more accurate constraints on position.

\subsection{Application to Data: Known and Unknown Sources }

As a test of this algorithm, we used HEAO1 ~A--2 MED data from a pointed
observation of the faintest test source, Abell~401. 
Shown in Figure~8 are position contours for both one orbit 
(about one hour; 352 time bins) of data,
and for the whole point (about three hours; 852 time bins).

\placefigure{figure8}

Finally, the full Bayes Poisson algorithm was used on the 
unidentified transient following GRB~780506. 
From comparing Figure~9  to Figure~7, one sees that
the position contours for the $\sim$50 minute emission shrank.
That should not be surprising.
The longer time--scale emission was not many $\sigma$ above background,
and by using the Poisson distribution,
one smoothly incorporates an implicit constraint that all the light--curve
parameters $\{s_k\}$ are non--negative. 
Notice that using the Poisson expression was important even when the total
average count--rate was above 30, when the source signal was faint
compared to background (Loredo\markcite{lo1} 1990).
(The contours visible in the upper left,
although formally allowed by the HED~1 data alone,
were ruled out after the fit by the lack of signal in the HED~2 data.)
Since this error--box has about an order of magnitude smaller area
than the previous error box,
it strengthened the supposition 
that this $\sim$50 minute X--ray glow came from the 
same source as the {\gray} burst (Connors and McConnell\markcite{co96} 1996).

\placefigure{figure9}

}


\begin{deluxetable}{cccccccc}
\tablecaption{Shape of the spectra for 7 time intervals \label{table1}}
\tablehead{
{} & {$\chi^2$} &
 \multicolumn{3}{c}{Black body component} & 
 \multicolumn{2} {c} {Power law component} &
 {$N_H$} \\
{} & {for 31} & 
 {$A_1$}&{Radius}&{kT}&
 {$A_2$} & {$\alpha$} &
 {$10^{23}$} \\
\colhead{No.} & \colhead{ DOF } & 
 \colhead{$\times 10^3$} & \colhead{km @ 1kpc}& \colhead{ keV } &
 \colhead{} & \colhead{} &
 \colhead{ cm$^{-2}$}
          }

\startdata
  1 & 26.9 &$.17\pm.01$&$.04\pm .004$ & $6.4\pm 2.0 $&$.47\pm .004$& $1.16\pm .01$ & $.3\pm .05$ \nl
  2 & 26.0 &$1.6\pm.1$&$.12\pm .01$ & $2.6\pm .05$&$.085\pm .005$ & $1.56\pm .03$ & $0.\pm .01$ \nl
  3 & 25.6 &$6.9\pm.4$&$.26\pm .03$ & $1.85\pm .03$&$.19\pm .006$ & $1.44\pm .01$ & $0.\pm .006$ \nl
  4 & 27.0 &$3.8\pm.4$&$.19\pm .04$ & $1.76\pm .05$&$.40\pm .01$ & $1.87\pm .02$ & $0.\pm .004$ \nl
  5 & 11.4 &$37.\pm6.$&$.60\pm 0.2$ & $0.85\pm .03$&$.24\pm .02$ & $2.34\pm .04$ & $0.\pm .008$ \nl
  6 & 13.3 &$120.\pm 100.$&$1.1\pm 0.9$ & $0.45\pm .09$&$.28\pm .03$ & $2.91\pm .07$ & $0.\pm .1$ \nl
  7 & 8.2  &$1\cdot 10^3\pm2 \cdot10^2$  &$3.1\pm 1.2$ & $0.42\pm .01$&$.093\pm .02$ & $2.21\pm .1$ & $0.\pm .1$ \nl
\enddata
\end{deluxetable}

{
\begin{table*}
\caption{Observations Used for Position--Finding Algorithms \label{table2}}
\begin{center}
\begin{tabular} {lcccccc}
\tableline
\tableline
{~} & 
{~} & {~} &
\multicolumn{2} {c} {Peak Source Rates} &
 \multicolumn{2} {c} {Average Background rates} 
 \\
{Source} & 
{D.O.Y.} & {No. of} &
 {Large FOV} & {Small FOV} &
 {Large FOV} & {Small FOV} 
 \\
{Name} & 
{1977 } & {5.12~s Bins} &
 {{${{ \rm cts}\over{ 5.12~s}}$}} & {{${{ \rm cts}\over{ 5.12~s}}$}} &
 {{${{ \rm cts}\over{ 5.12~s}}$}} & {{${{ \rm cts}\over{ 5.12~s}}$}}
 \\
\tableline
Crab\tablenotemark{s,a}     & 260 & 182 & 2160 & 1743 &$23.3\pm 1.5$ & $13.4\pm 1.2$ \nl
GX301-02\tablenotemark{p,a} & 390 & 632 &  114 &   80 &$23.6\pm 1.5$ & $13.6\pm 1.2$  \nl
Abell~401\tablenotemark{p,b}& 404 & 832 &   49 &   35 &$23.9\pm 1.5$ & $13.6\pm 1.2$ \nl
Abell~401\tablenotemark{p,b}& 404 & 352 &   49 &   35 &$23.9\pm 1.5$ & $13.6\pm 1.2$ \nl
GRB~780506\tablenotemark{p,c}& 491 & 24 & 1386 & 1012 &$56.2\pm 0.4$ &$30.1\pm 0.3$ \nl
Afterglow\tablenotemark{p,c}& 491 & 464 &   88 &   51 &$56.2\pm 0.4$ &$30.1\pm 0.3$ \nl
\tableline
\end{tabular}
\end{center}

\tablenotetext{s}{Scanning data}
\tablenotetext{p}{Pointed data}
\tablenotetext{a}{Source position from Bradt and McClintock (1984).}
\tablenotetext{b}{Source position from SIMBAD.}
\tablenotetext{c}{Source position from this work.}

\tablenum{2}

\end{table*}

}
%


\newpage


\figcaption[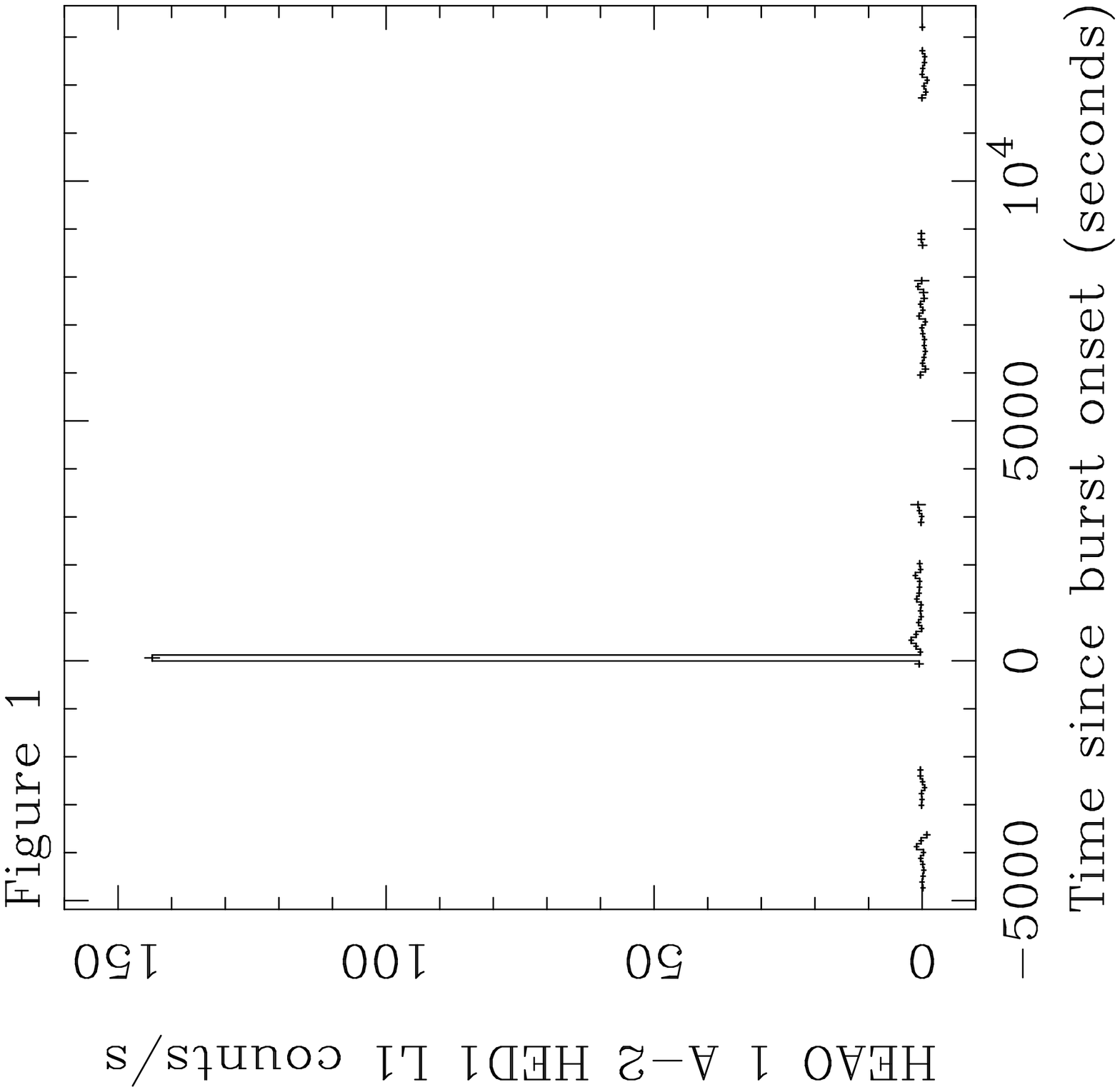] 
{{\it HEAO~1} A--2 HED1 Layer~1 2--20 keV 
background--subtracted cts--s$^{-1}$ versus time since burst onset 
(at 82632.97~s U.T.)
during the 6 hr point on May 6, 1978.
The large gaps are due to Earth occultation; the smaller ones to
times of higher particle background. 
 \label{figure1}}


\figcaption[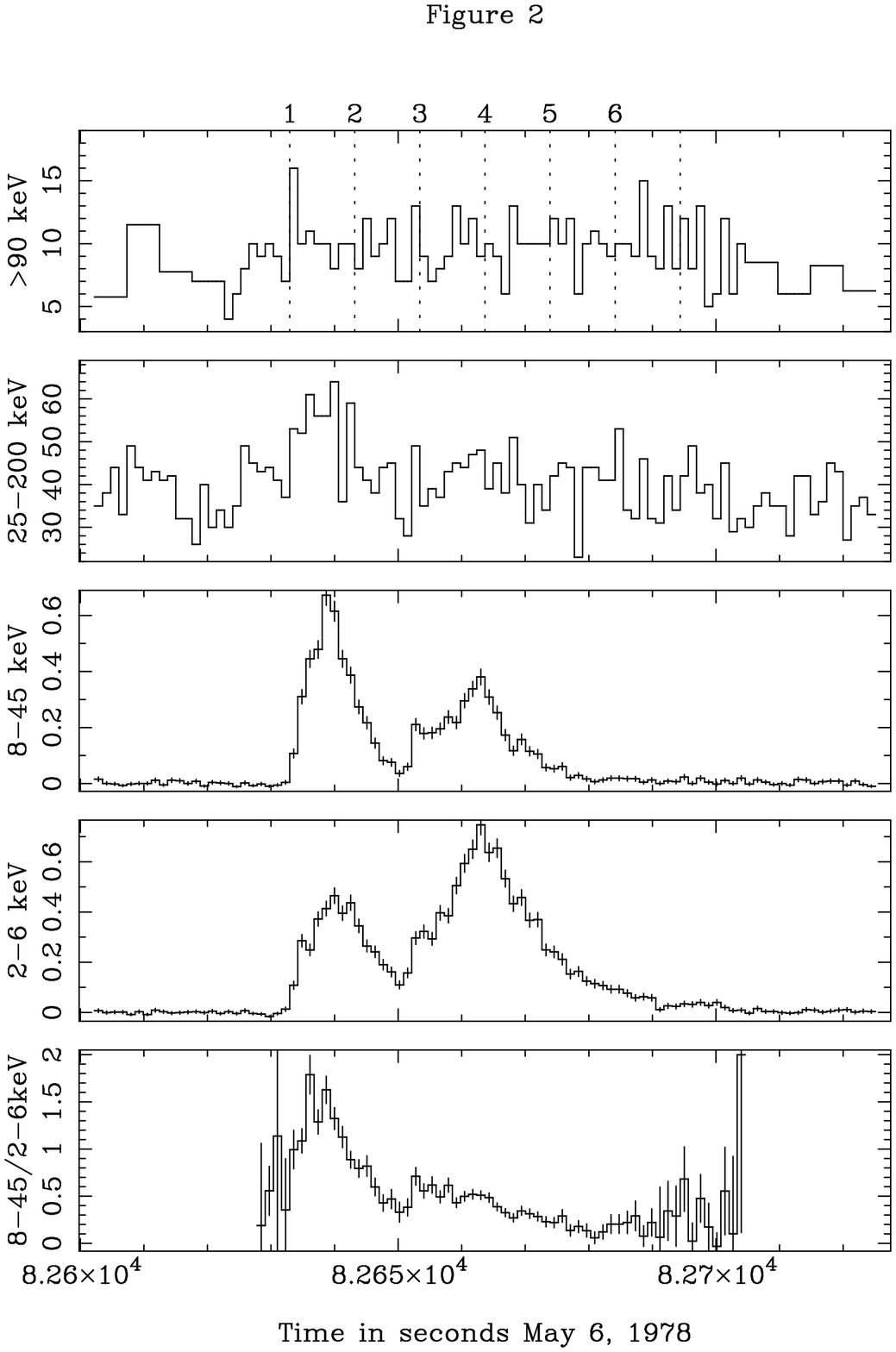] 
{{\it HEAO~1} A--4 {\it(top two panels}; in cts/1.28~s) and 
A--2 {\it(middle two panels}; in cts--cm$^{-2}$--s$^{-1}$) 
light--curves,
and 8-45~keV / 2--6~keV hardness ratio {\it(bottom panel)}
during the burst, in 1.28~s bins.  
At the top of the figure we have marked the 10.24~s spectra accumulation
intervals.
\label{figure2}}


\figcaption[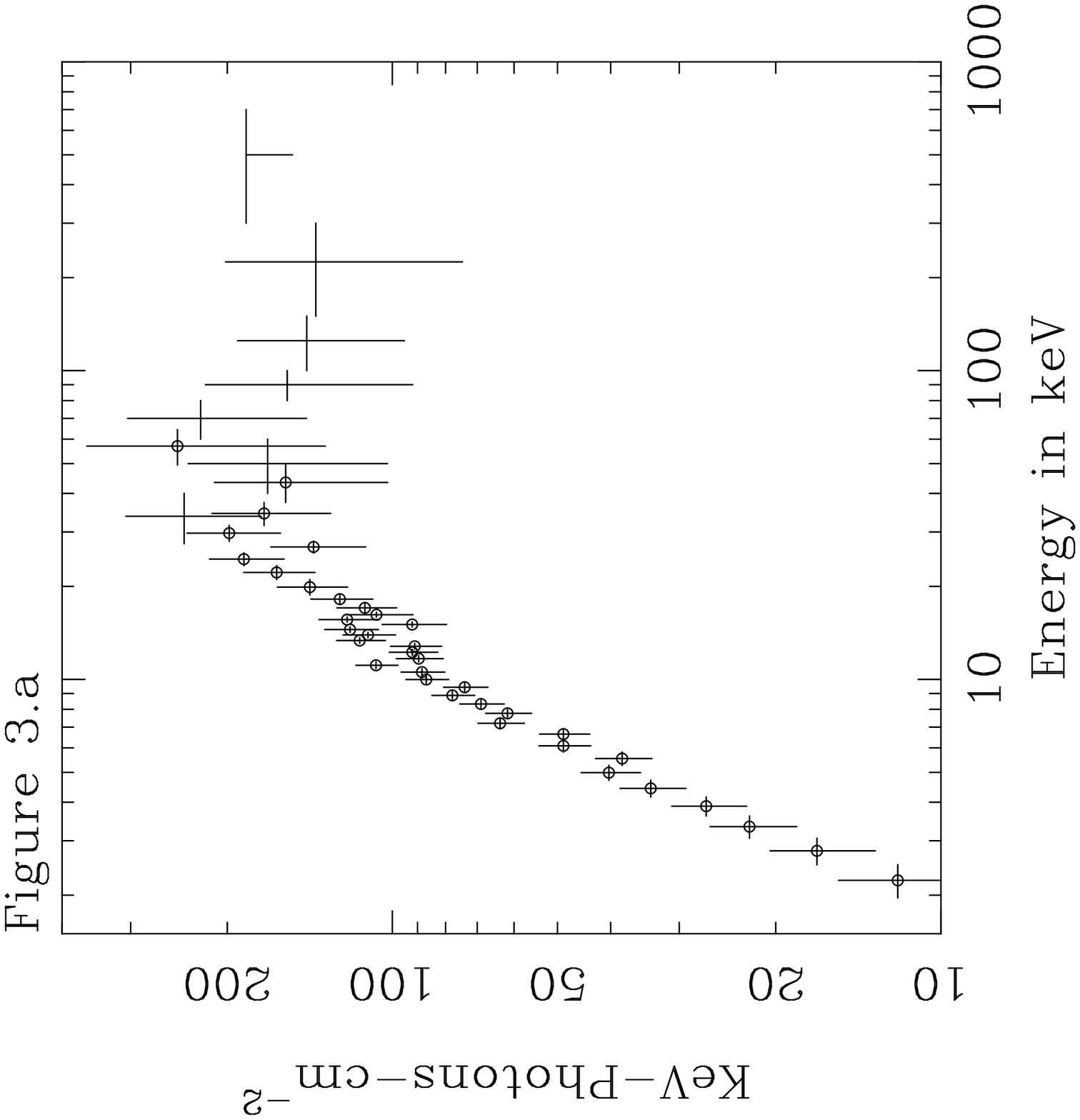,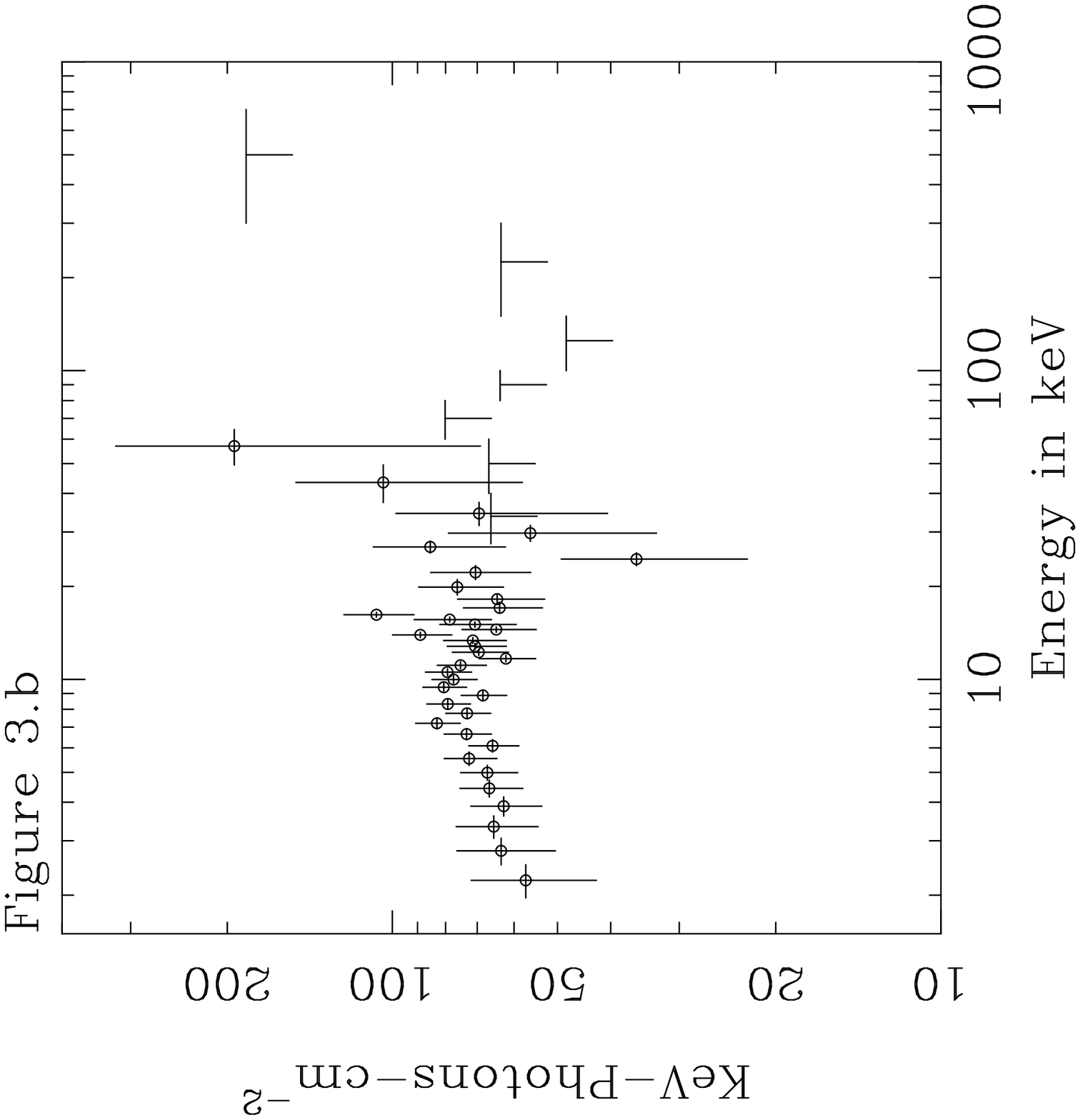,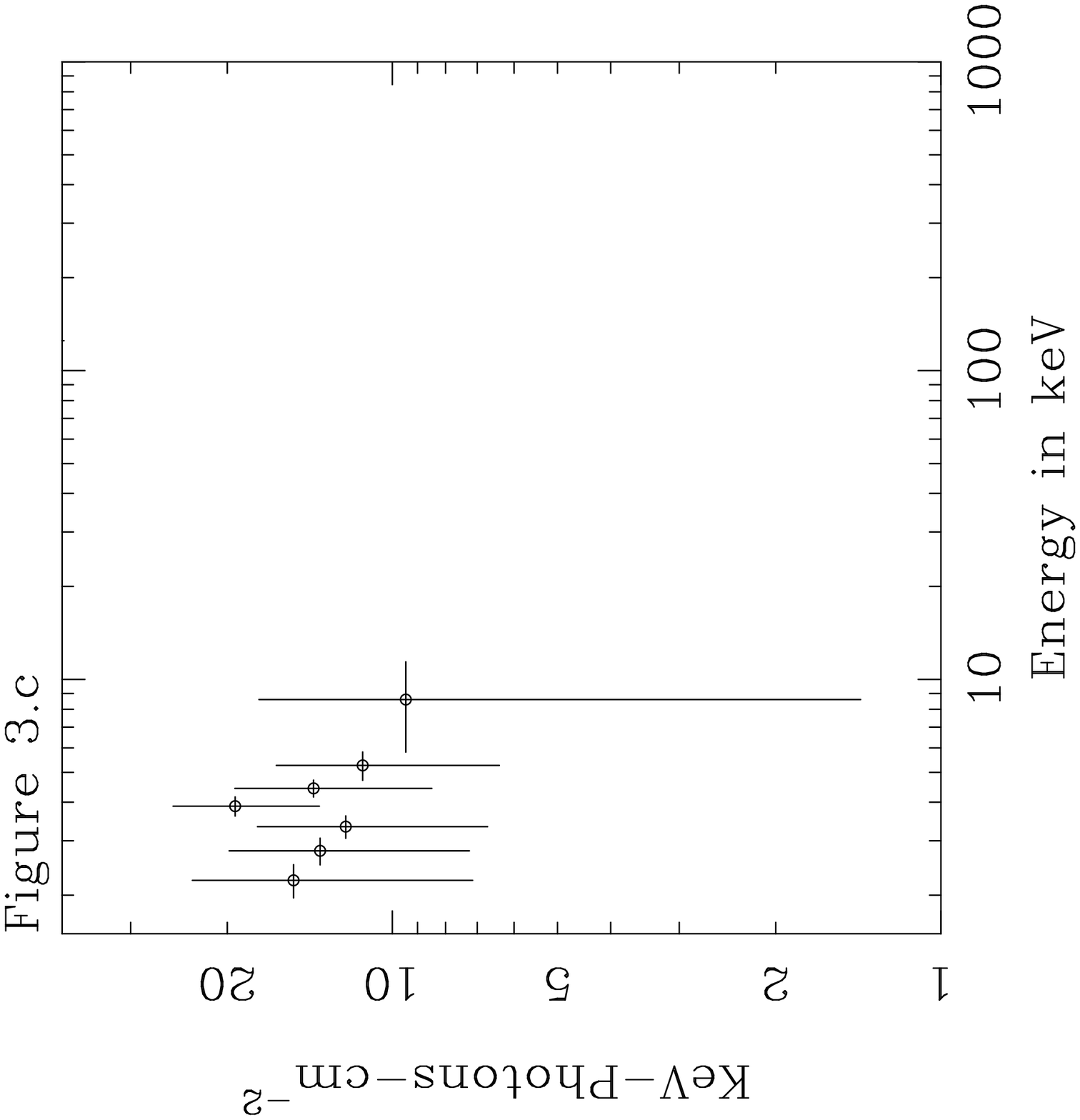] 
{{\it HEAO~1} A--2 (open circles) and A--4 $\nu F_{\nu}$ spectra, 
in keV--photons--cm$^{-2}$
 for 
(a) the first peak of the burst, intervals 1+2;
(b) the second peak, intervals 3+4+5;
(c) the afterglow.
\label{figure3}}


\figcaption[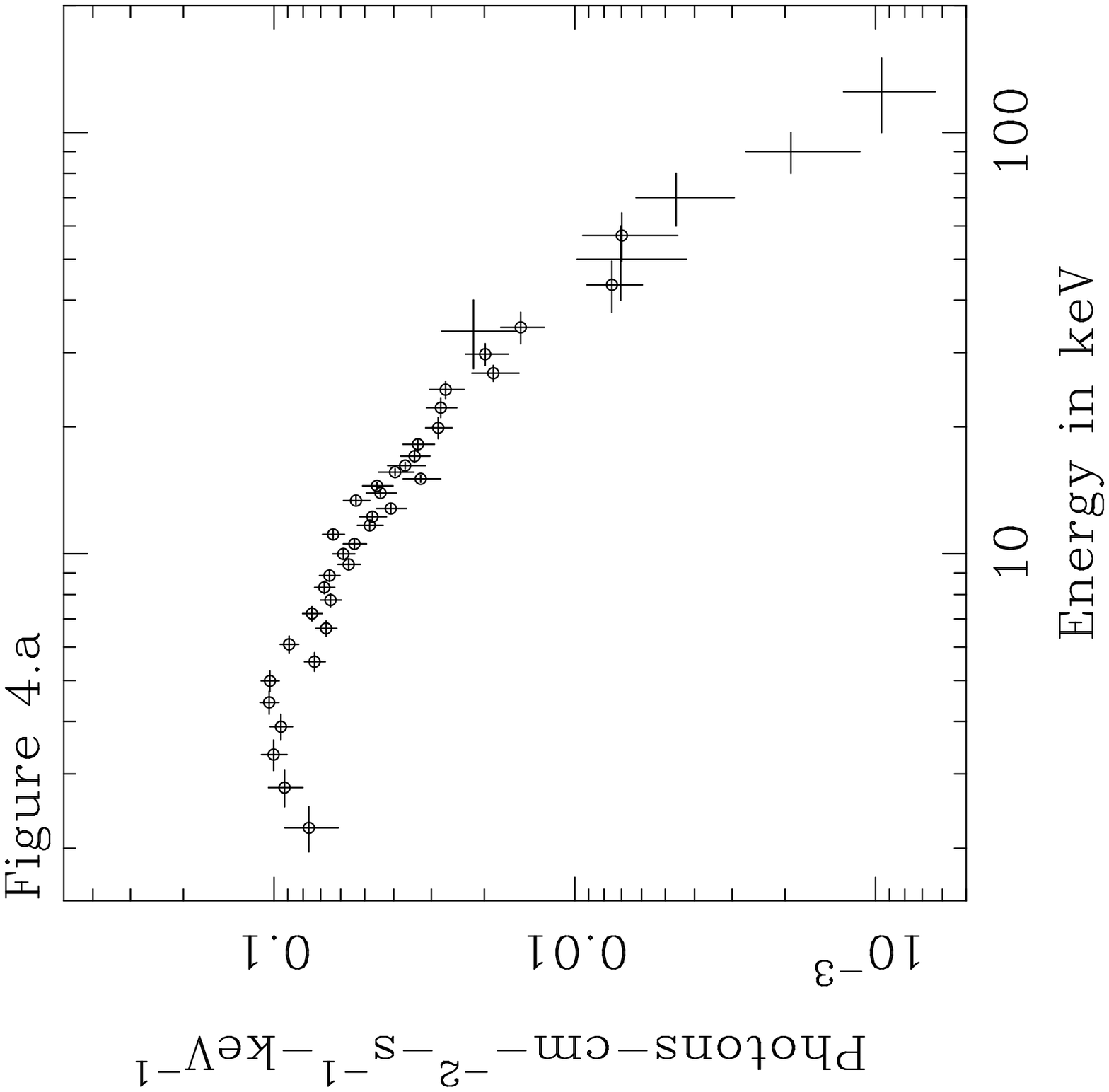,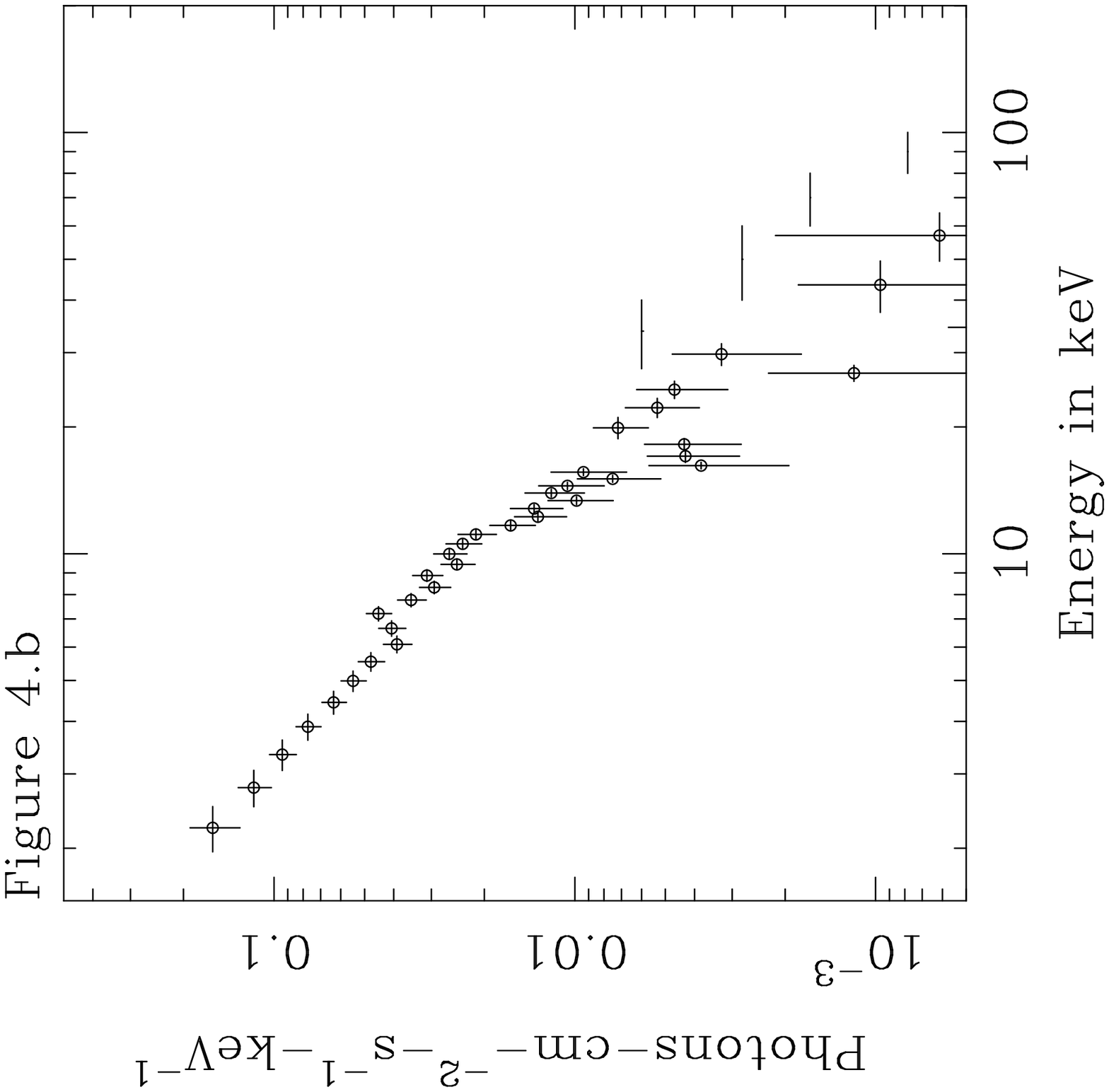,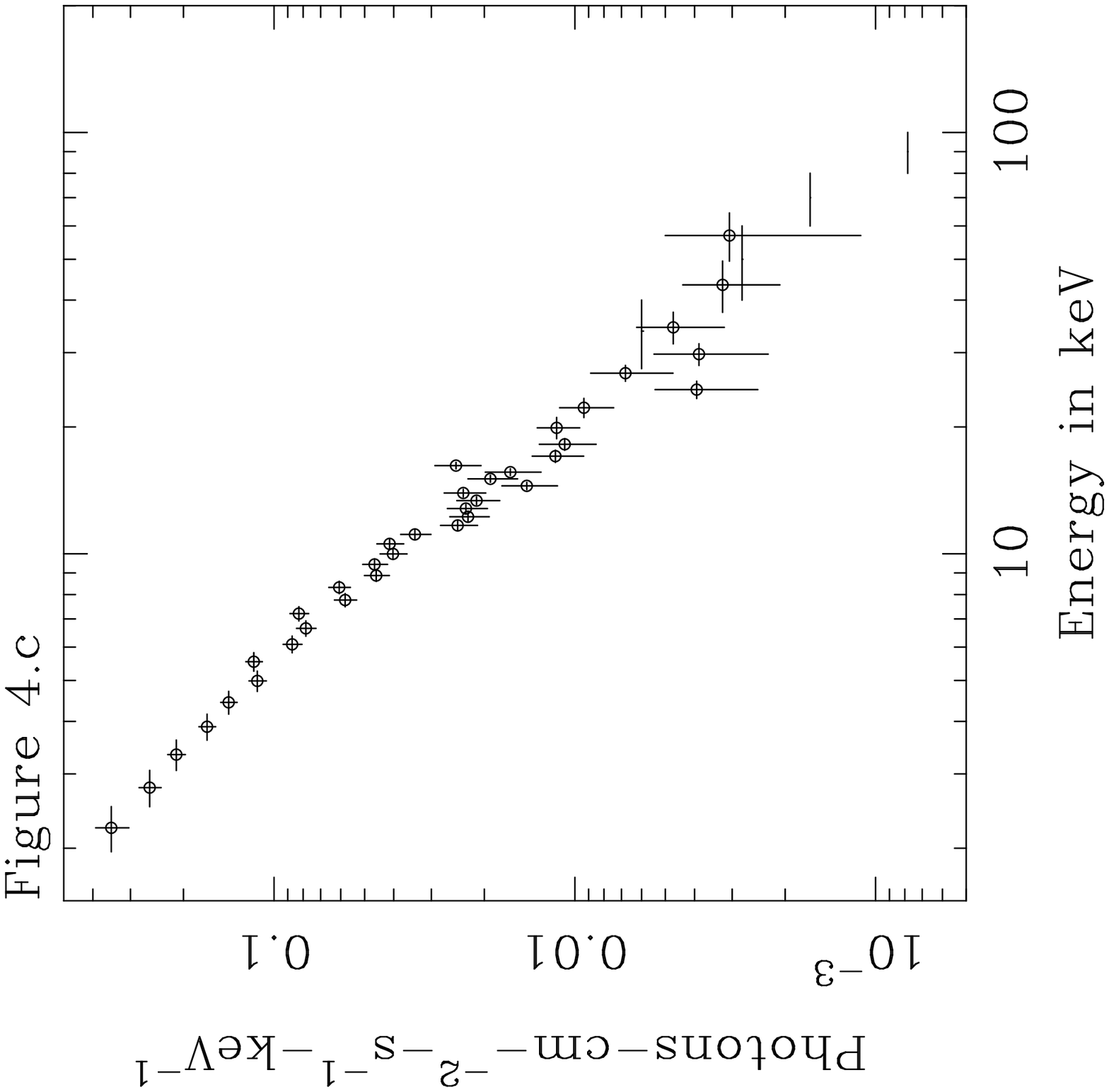,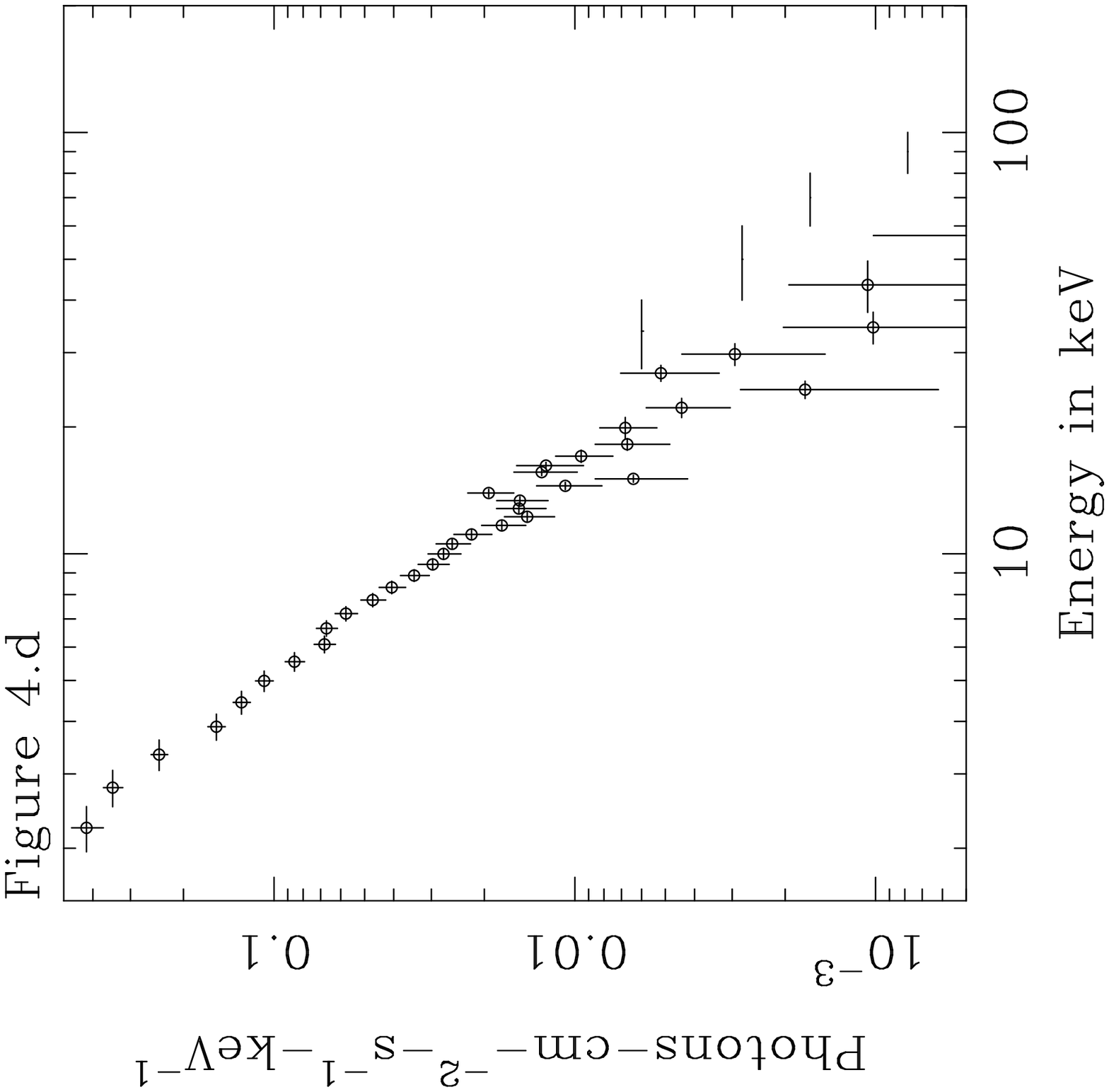,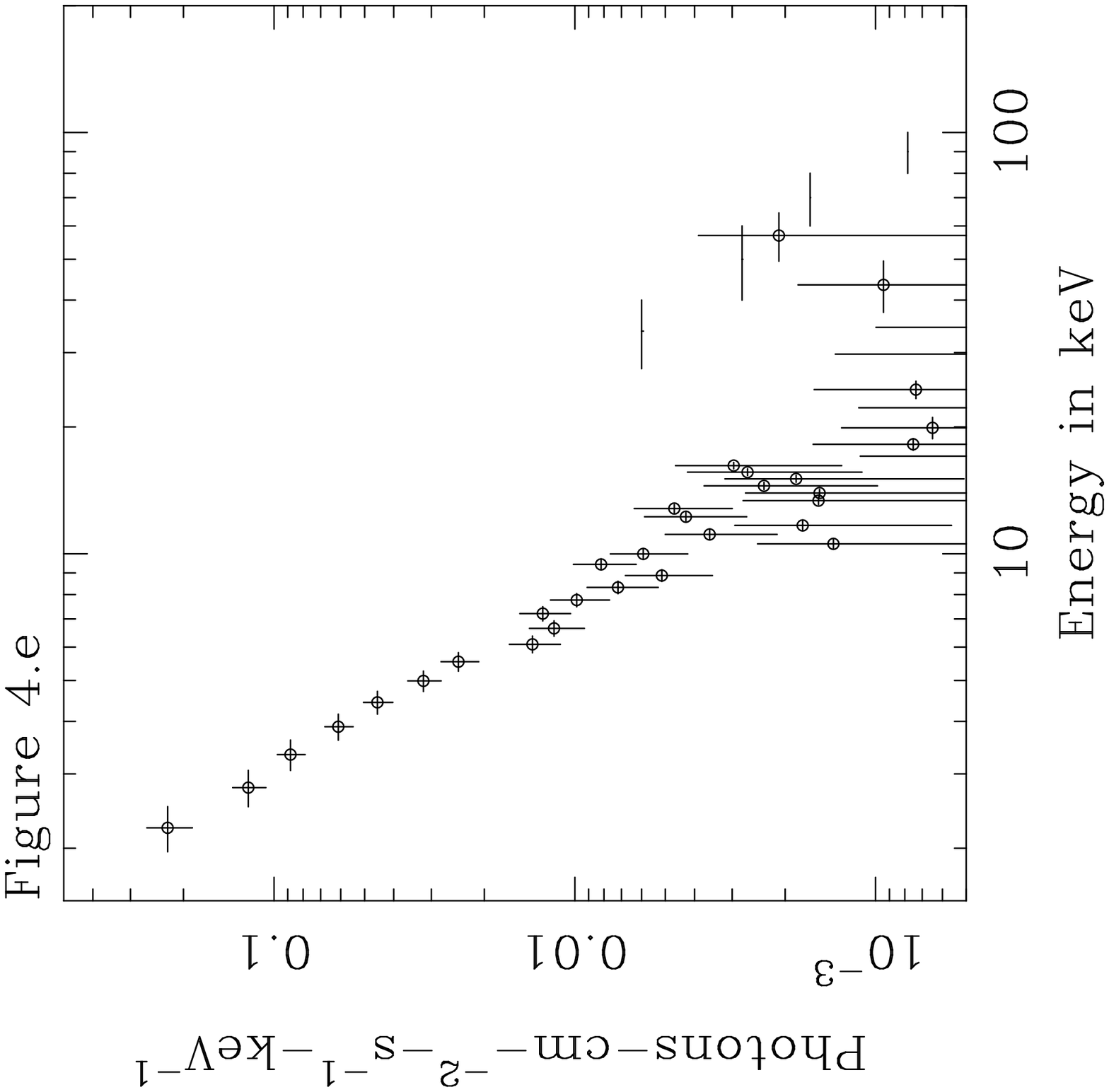,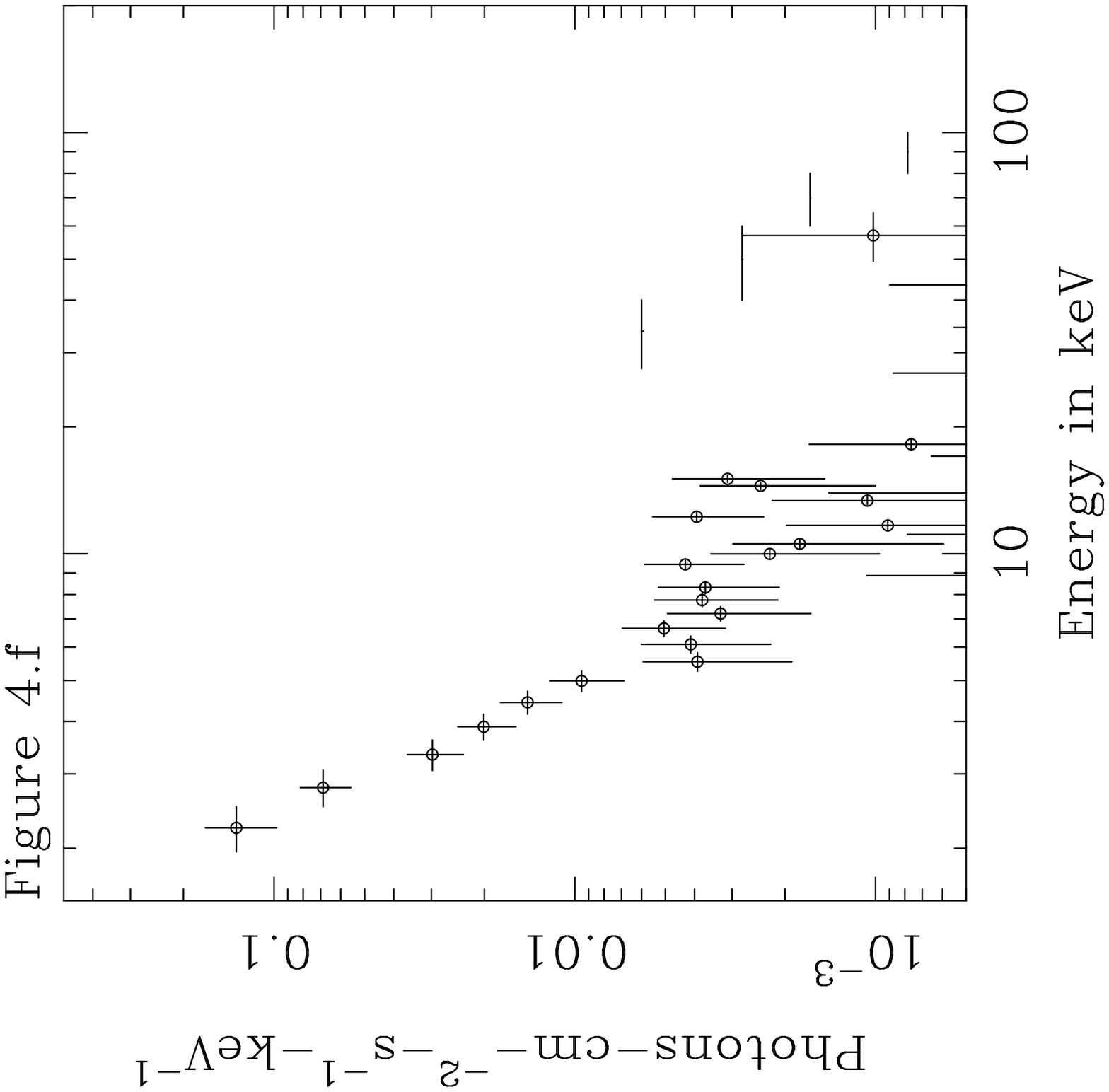,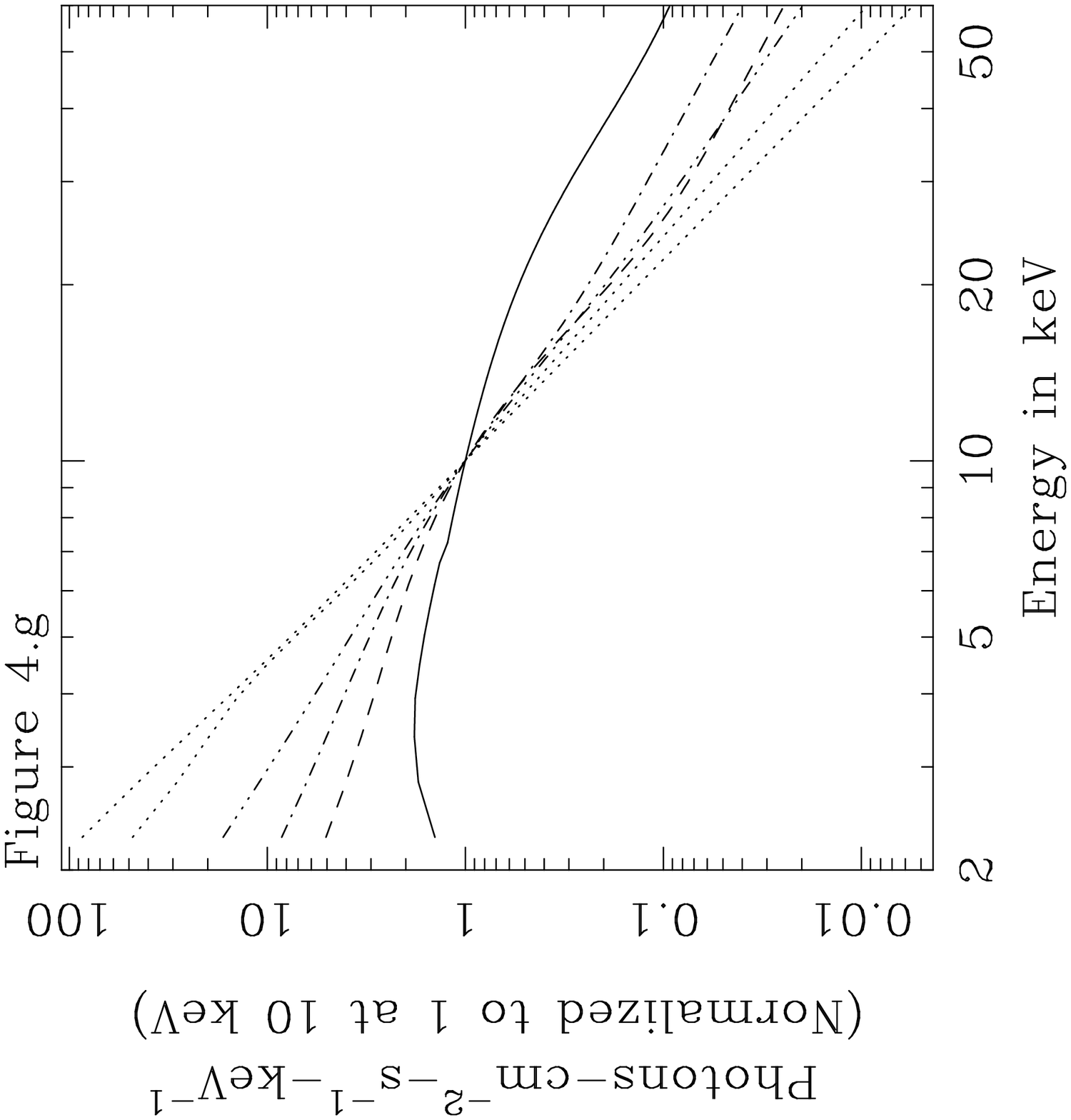] 
{ {\it (a)--(f) HEAO~1} A--2 {\it (open circles)} and A--4 spectra, 
in photons--cm$^{-2}$--s$^{-1}$--keV$^{-1}$
 for each of the six accumulation intervals marked in Figure 3.
Except for the first spectrum (Figure 4.a), the {\it HEAO~1} A--4
data are upper limits.
{\it (g)} Best--fit spectra for each time interval: 1{\it (solid line)}, 
2 {\it(dashed line)}, 3 {\it(dot--dashed line)},  
4 {\it (dash--double--dotted line)}, and 5 and 6 {\it(dottted lines)}.
\label{figure4}}


\figcaption[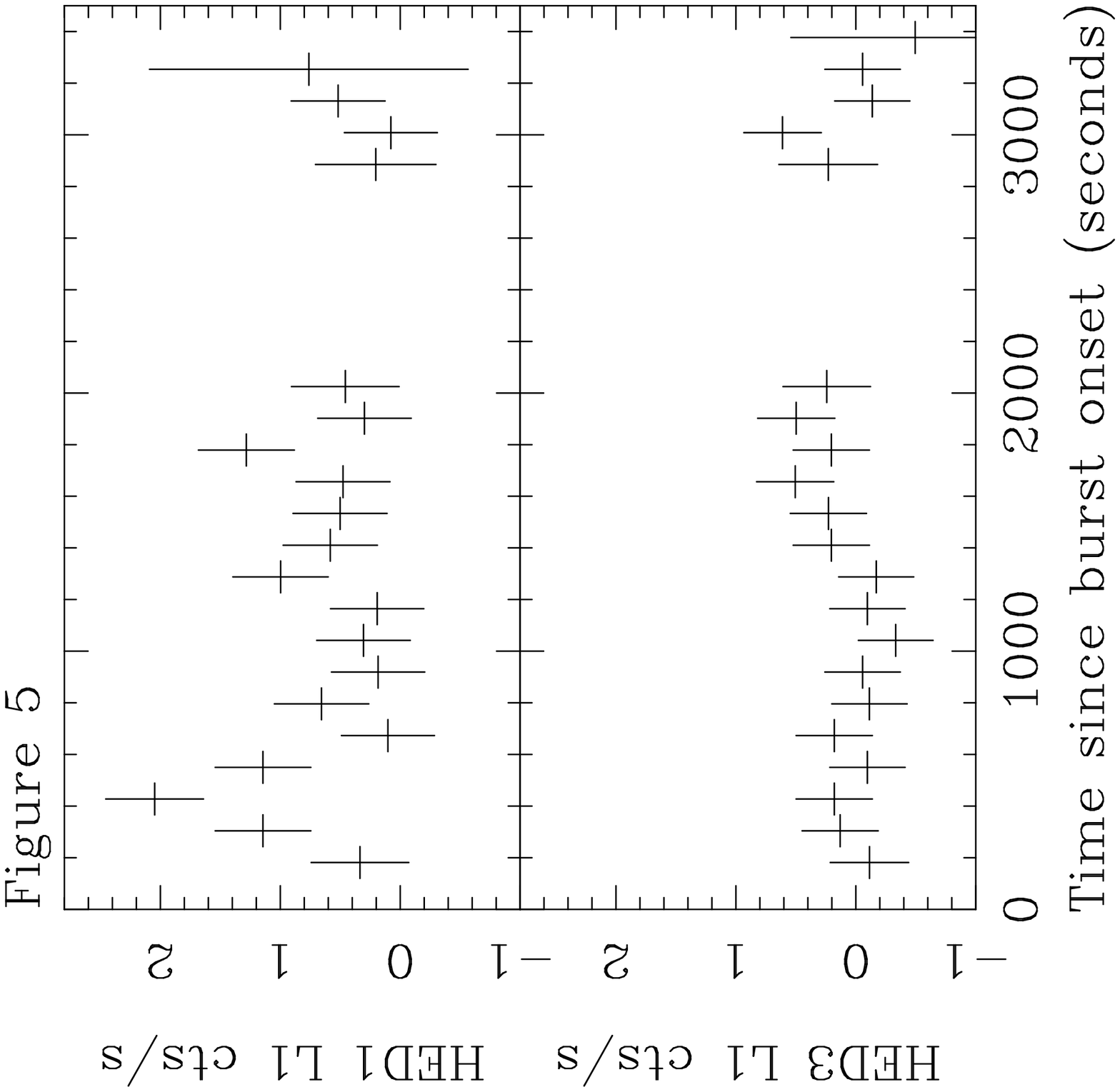] 
{Background--subtracted 2--20 keV cts--s$^{-1}$ from
{\it HEAO~1} A--2 HED1 Layer~1 {\it (upper panel)}
and {\it HEAO~1} A--2 HED3 Layer~1 {\it (lower panel)} 
showing the faint afterglow 
immediately following the burst.
(The burst itself is off-scale.)
\label{figure5}}

\figcaption[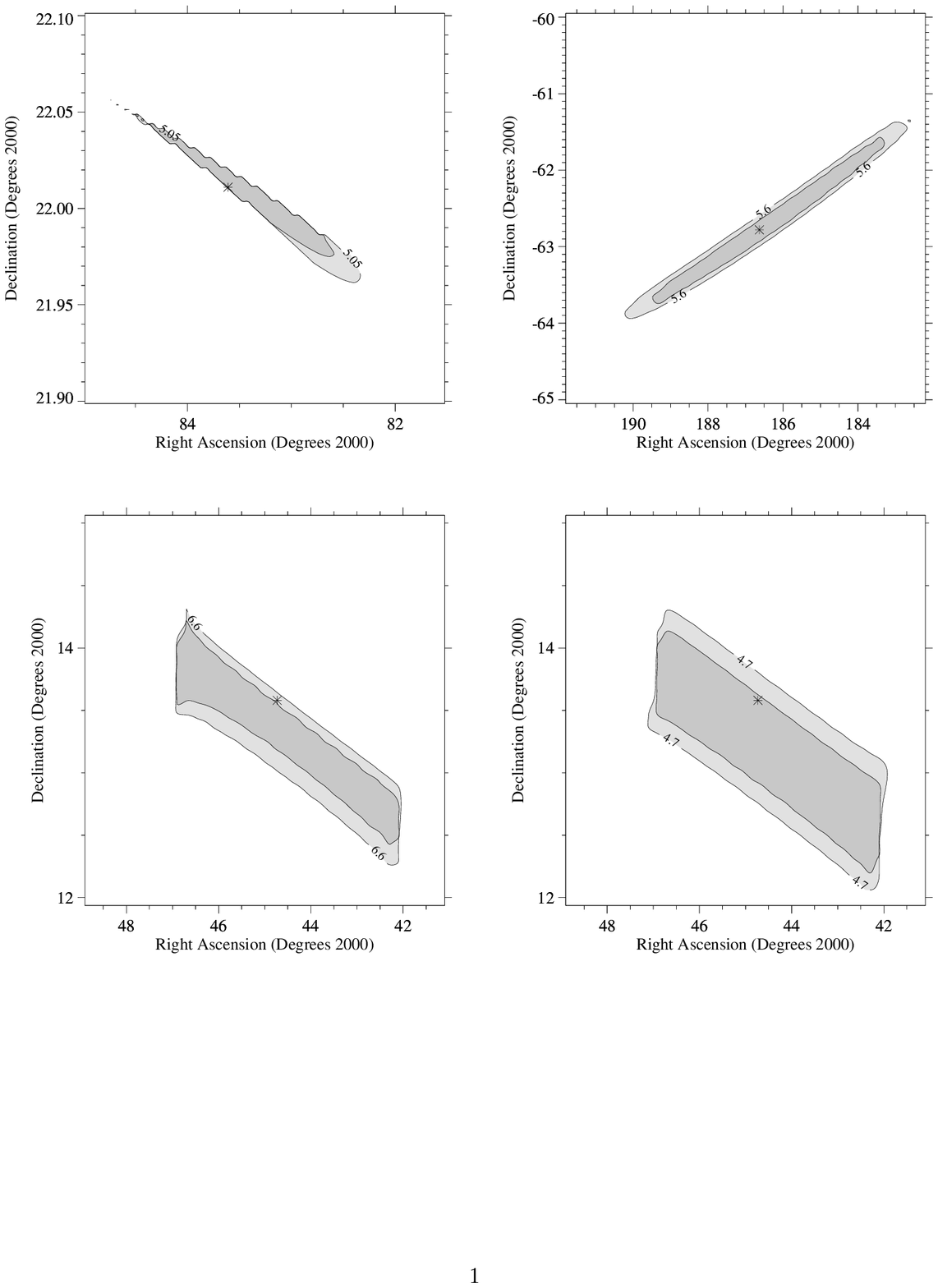]
{ {\it Upper panels}:
95.45\% and 99.73\% (equivalent to 2 and 3 $\sigma$) credible regions
 for 5 scans
of Crab data {\it (left panel)};  
and 6 hr of pointed data on the highly variable
source GX~301-02 {\it (right panel)}.
{\it Lower panels}:
95.45\% and 99.73\% (equivalent to 2 and 3 $\sigma$) credible regions
for 3 hr {\it (left panel)}
and 1 orbit, or about 1 hr {\it (right panel)} of pointed data
on the fainter steady source Abell~401.
\label{figure6}}


\figcaption[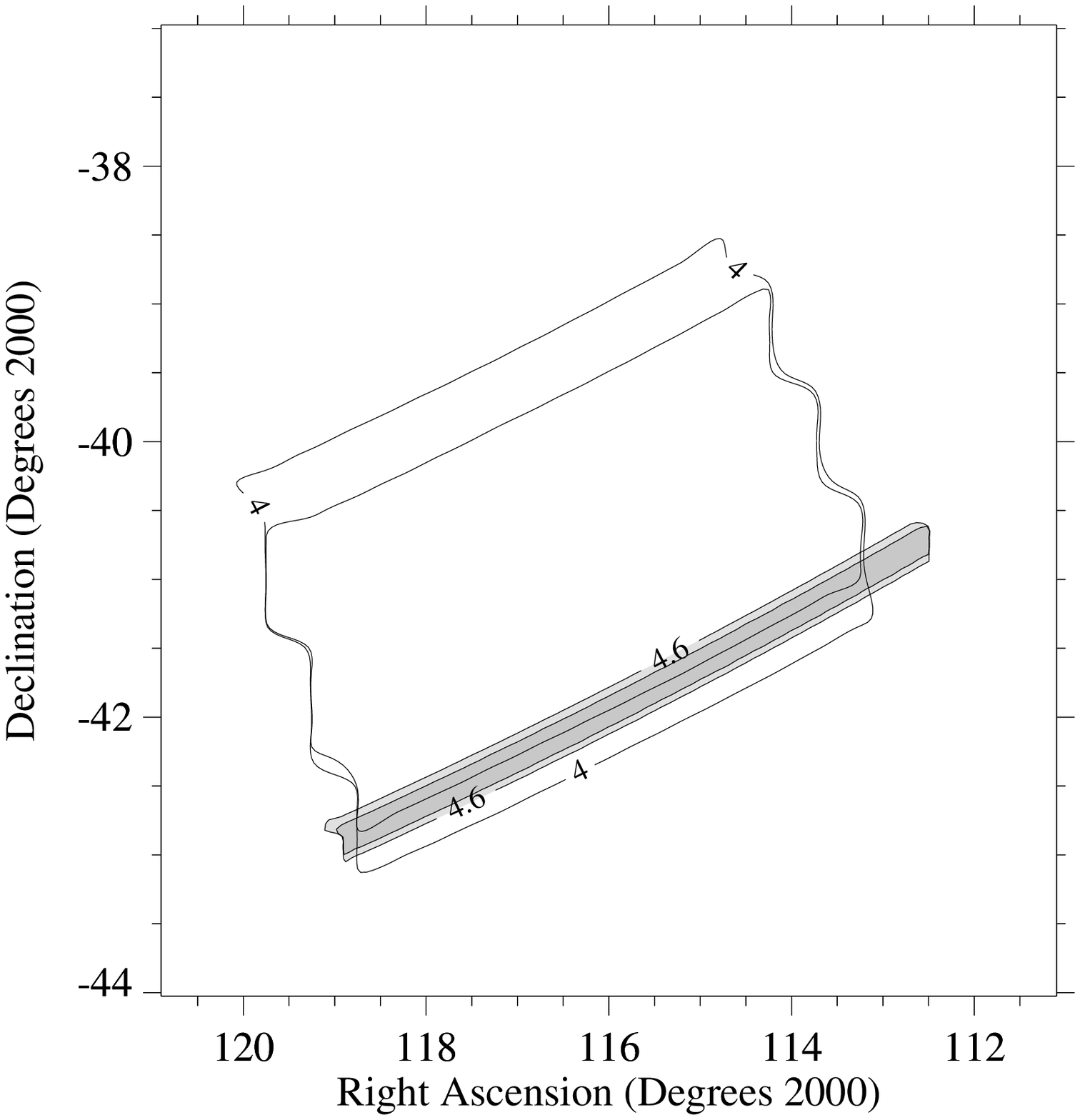]
{95.45\% and 99.73\% (equivalent to 2 and 3 $\sigma$) credible regions
for GRB~780506 {\it(grayscale contours)}
plus extended X--ray emission {\it(line contours)}, 
for data from the (90 minute) orbit containing the burst.
Both were calculated using the Bayesian likelihood ratio
derived assuming a Gaussian approximation for the direct probability,
appropriate for high signal--to--noise data.
\label{figure7}}


\figcaption[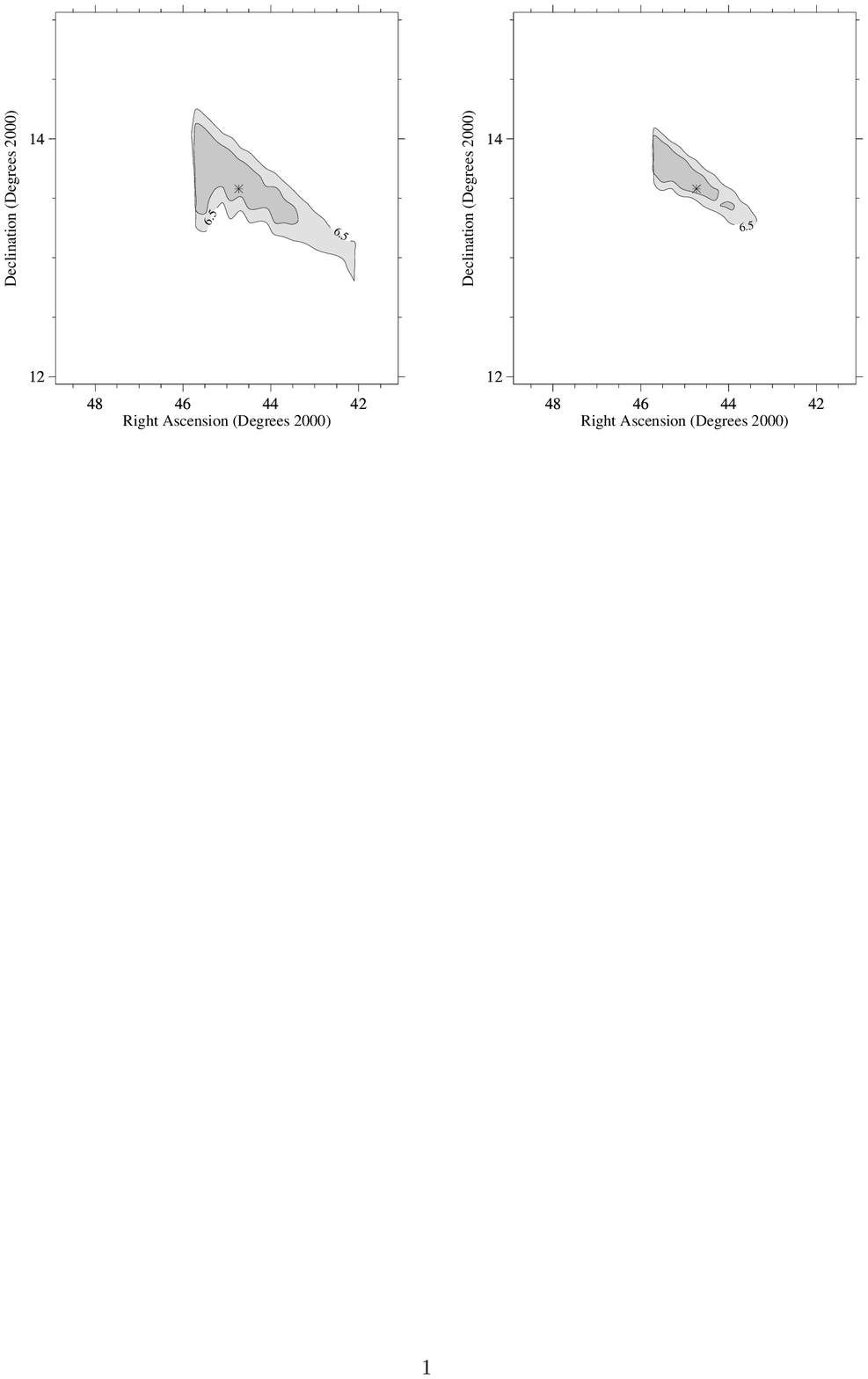]
{{\it (Left panel)}
95.45\% and 99.73\% (equivalent to 2 and 3 $\sigma$) credible regions
for 1 orbit (about 1 hr)
of {\it HEAO~1} A--2 MED data of Abell~401.	
{\it (Right panel)} the same for the full point (about 3 hr of data).
\label{figure8}}


\figcaption[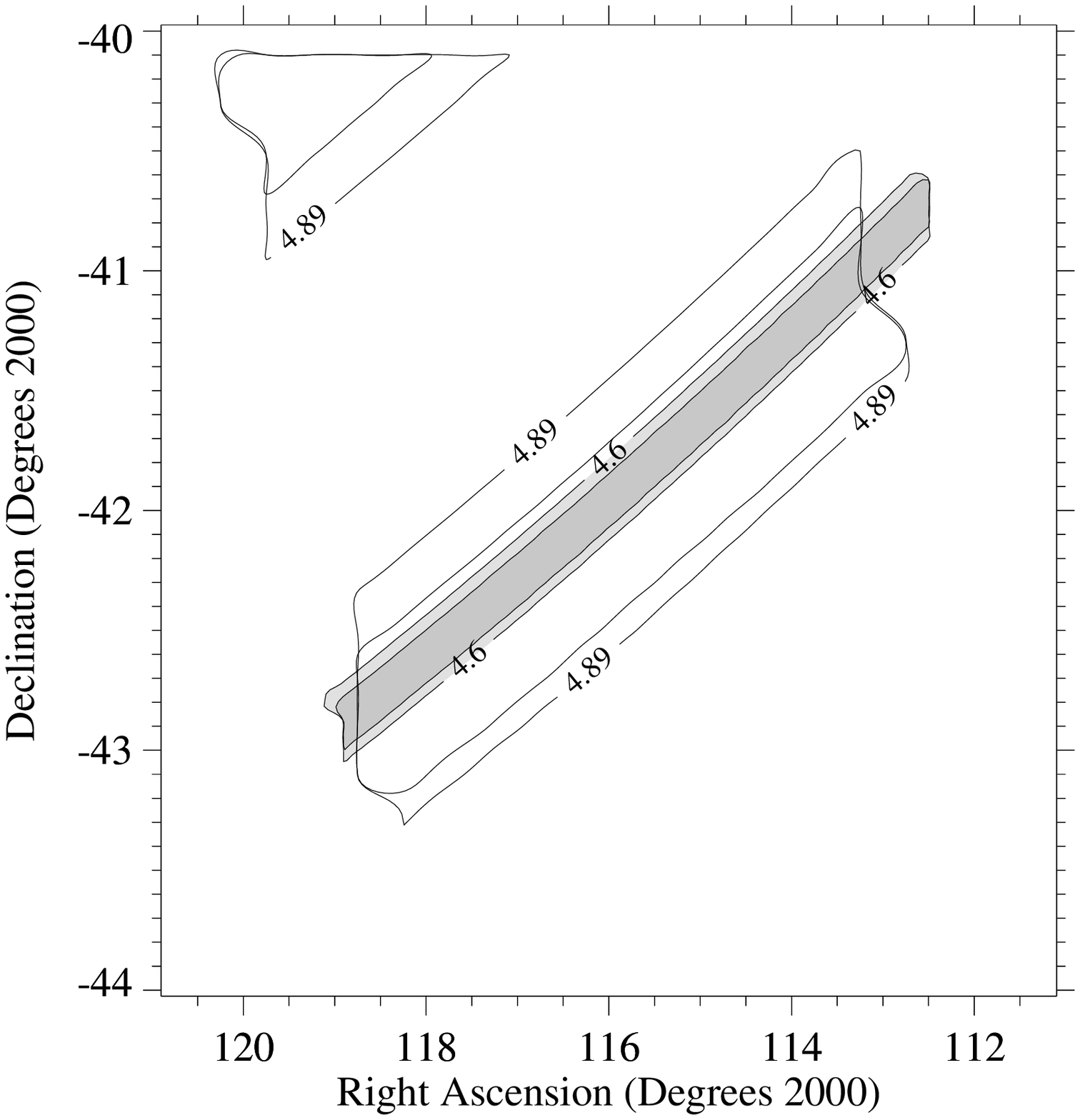]
{95.45\% and 99.73\% (equivalent to 2 and 3 $\sigma$) credible regions
for GRB~780506 {\it (grayscale contours)}
plus extended X--ray emission {\it (line contours)}, 
for data from the (90 minute) orbit containing the burst.
The latter was calculated using the Bayesian likelihood ratio
derived from the full Poisson expression for direct probability,
rather than a Gaussian approximation.
\label{figure9}}



\clearpage

\plotone{f1.ps}

\clearpage

\plotone{f2.ps}

\clearpage

\plotone{f3a.ps}

\clearpage

\plotone{f3b.ps}

\clearpage

\plotone{f3c.ps}

\clearpage

\plotone{f4a.ps}

\clearpage

\plotone{f4b.ps}

\clearpage

\plotone{f4c.ps}

\clearpage

\plotone{f4d.ps}

\clearpage

\plotone{f4e.ps}

\clearpage

\plotone{f4f.ps}

\clearpage

\plotone{f4g.ps}

\clearpage

\plotone{f5.ps}

\clearpage

\plotone{f6.ps}

\clearpage

\plotone{f7.ps}

\clearpage

\plotone{f8.ps}

\clearpage

\plotone{f9.ps}


\newpage

\begin{references}

\reference{ab1} Abramowitz, M. \& Stegun, I., 1972, {\it Handbook of
Mathematical Functions}, (New York: Dover Publications, Inc.) \par

\reference{ba1} Band, D., et al. 1992, in {\it Gamma--Ray
Bursts}, AIP Conf. Proc. 265, eds. W. S. Paciesas \& G. B. Fishman,
(New York: AIP Press), p 169 \par

\reference{ba2} Band, D., et al. 1993, {\apj}, 413, 281 \par

\reference{boella} Boella, G., et al., 1997, {A\&AS}, 122, 299 \par

\reference{be97} Belli, B. M., 1997, \apjl, 479, L31 \par

\reference{be1} Bevington, P. R., 1969, {\it Data Reduction and
Error Analysis for the Physical Sciences}.  (New York: McGraw-Hill) \par

\reference{bradtm} Bradt, H. V. D., \& McClintock, J. E. 1984, {ARA\&
A}, 21, 13 \par

\reference{bra94} Brainerd, J. J., 1994, \apj, 428, 21 \par

\reference{ca1} Cash, W., 1978, \apj, 228, 939 \par


\reference{ch1} Chernenko, A. M. \& Mitrofanov, I. G., 1994, 
{\it Gamma--Ray Bursts}, AIP Conference Proc. No 307,
ed G. J. Fishman, J. J. Brainerd, \& K. Hurley (New York:  AIP) 293 \par

\reference{chernenko98} Chernenko, A. M., et al., 1998, in
{\it Proceedings of the 4th Huntsville Symposium on Gamma--ray Bursts}
AIP Conference Proc. 428, Meegan, C. A., Preece, R. D. \& Koshut, T. M.
(New York: AIP), 30 \par

\reference{ch2} Chipman, E., 1994, 
{\it Gamma--Ray Bursts}, AIP Conference Proc. No 307,
ed G. J. Fishman, J. J. Brainerd, \& K. Hurley (New York:  AIP)  202 \par

\reference{cl1} Cline, T. L., et al. 1979, \apjl, 232, L1 \par

\reference{cohen97} Cohen et al. 1997, \apj, 488, 330 \par

\reference{co1} Connors, A. 1988,
Ph.D. thesis, University of Maryland \par

\reference{cosesw86} Connors, A., Serlemitsos, P. J., \& Swank, J. H., 1986,
\apj,  303, 769 \par

\reference{co95}Connors, A. \& McConnell, M. 1995, in Proc. 24th
Int. Cosmic-Ray Conf., Rome, Italy, 2, 57 \par


\reference{co96} Connors, A. \& McConnell, M. M., 1996, in 
{\it Proc. of the 3rd Huntsville Symposium on Gamma--Ray Bursts
AIP Conf. Proc. 384},
ed. C. Kouveliotou, M. F. Briggs, \& G. J. Fishman,
(New York: AIP) 607 \par

\reference{co97a} Costa, E., et al., 1997, Nature, 387, 783 \par

\reference{co97b} Costa, E., et al., 1998, in
{\it Proceedings of the 4th Huntsville Symposium on Gamma--ray Bursts}
AIP Conference Proc. 428, Meegan, C. A., Preece, R. D. \& Koshut, T. M.
(New York: AIP), 409 \par

\reference{di1} Dingus, B., et al., 1994, 
{\it Gamma--Ray Bursts}, AIP Conference Proc. No 307,
ed G. J. Fishman, J. J. Brainerd, \& K. Hurley (New York:  AIP) 22 \par

\reference{ep1} Epstein, R.~I., 1986, in 
{\it Radiation Hydrodynamics in Stars and
Compact Objects}, Proceeding of IAU Colloquium 89, (Spring
Verlag: New York) eds.~ K.-H. Winkler \& D. Mihalas) 305 \par


\reference{fe1} Fenimore, E. E. et al., 1988, \apj, 335, L71 \par

\reference{fe2} Fenimore, E. E., in 't Zand, J.J.M., 
Norris, J.P., Bonell, J. T., \& Nemiroff, R. J. 1995,
\apjl, 488, L101 \par

\reference{fi1} Fishman, G. et al., 1994, \apjs, 92, 229 \par

\reference{fo1} Ford, L., Band, D., Matteson, J., Teegarden, B. \&
Paciesas, W. 1994, {\it Gamma--Ray Bursts}, AIP Conference Proc. No 307,
ed G. J. Fishman, J. J. Brainerd, \& K. Hurley (New York:  AIP) 298 \par

\reference{fo2} Forrest, D., et al., 1995, \apss, 231, 459 \par

\reference{gi1} Gilman, D., Metzger, A. E., Parker, L. H., Evans, L. G., \&
Trombka, J. I., 1980, \apj, 236, 951 \par

\reference{gr1} Graziani, C., Fenimore, E. E., Murakami, T., Yoshida, A.,
Lamb, D. Q., Wang, J. C. L., \& Loredo, T. J., 
1992, in {\it Gamma--Ray Bursts: Observations, Analyses and Theories}, eds. 
C. Ho, R. I. Epstein, \& E. E. Fenimore
(Cambridge: Cambridge University Press), 407 \par

\reference{gl92} Gregory, P. C., \& Loredo, T. J. 1992, {\apj} 398, 146 (GL92)

\reference{halpern} Halpern, J., et al., 1997, IAU Circ. 6788.\par

\reference{heise} Heise, J., et al.,  1998, in 
{\it Proceedings of the 4th Huntsville Symposium on Gamma--ray Bursts}
AIP Conference Proc. 428, Meegan, C. A., Preece, R. D. \& Koshut, T. M.
(New York: AIP), 397 \par

\reference{heise97b} Heise, J., et al., 1997, IAU Circ. 6787 \par

\reference{hu1} Hueter, G., 1987,
Ph.D. thesis, University of California at San Diego \par

\reference{hu2} Hurley, K., {\it et. al.}, 1994 \nat, { 372}, 652 \par

\reference{im1}  Imamura, J.~N.~and Epstein, R.~I.~, 1987 \apj, 313, 711 \par

\reference{ka1} Kargatis, V.E., Liang, E.P., Hurley, K. C.,
Barat, C., Evans, E., \& Niel, M., 1994, \apj, 422, 260 \par

\reference{ka2} Katoh, M., Murakami, T., Nishimura, J., Yamagami, T.,
Tanaka, Y., \& Tsunemi, H., 1984, in
{\it High Energy Transients in Astrophysics}, AIP Conference Proc. No 115,
ed S. E. Woosley (New York:  AIP), 390. \par

\reference{Katz97} Katz, J., 1997, \apj, 490, 633.\par

\reference{kapisa97} Katz, J. I., Piran, T., \& Sari, R., 1997, 
submitted to Phys Rev. B. \par

\reference{ke1} Kerr, F. J., Bowers, P. F., Jackson, P. D., \& Kerr, M.
 1986, A\&ASS, 66, 373 \par

\reference{ki1} Kippen, R. M., et al., 1997, Adv. Space Res., in press \par

\reference{kl1} Klebesadel, R., Laros, J., \& Fenimore, E., 1984, \baas,
{ 16}, 1016 \par

\reference{kl2} Klebesadel, R. 1992, in 
{\it Gamma--Ray Bursts: Observations, Analyses and Theories}
ed. C.~Ho, R.~I.~Epstein \& E.~E.~Fenimore 
(Cambridge: Cambridge University Press) 161 \par

\reference{ko1} Kouveliotou, C., Meegan, C. A., Fishman, G. J., Bhat, P. N.,
Briggs, M. S., Koshut, T. M., Paciesas, W. S., \& Pendleton, G. N., 1994, in
{\it Gamma--Ray Bursts}, AIP Conference Proc. No 307,
ed G. J. Fishman, J. J. Brainerd, \& K. Hurley (New York:  AIP) 167 \par

\reference{koetal96}   Kouveliotou, C., et al. 1996, in 
{\it Proc. of the 3rd Huntsville Symposium on Gamma--Ray Bursts
AIP Conf. Proc. 384},
ed. C. Kouveliotou, M. F. Briggs, \& G. J. Fishman,
(New York: AIP) 42 \par


\reference{ln1} Lang, K., 1980, {\it Astrophysical Formulae}
(New York:  Springer--Verlag) \par

\reference{la1} Laros et al., 1984, in
{\it High Energy Transients in Astrophysics} ed. S. E. Woosley 
(New York: AIP) 378 

\reference{la2} Laros et al., 1984, ApJ, 286, 681 \par

\reference{la3} Laros, J. \& Nishimura, J., 1986, 
in {\it Gamma--Ray Bursts}, AIP Conference Proceedings No 141, eds.
E. P. Liang \& V. Petrosian, (New York: AIP) 79 \par

\reference{li1} Liang, E. P., 1994,
in {\it Gamma--Ray Bursts}, AIP Conference Proceedings No 307, eds.
G. J. Fishman, J. J. Brainerd, \& K. Hurley, (New York: AIP) 351 \par

\reference{li97} Liang, E. P., Kusunose, M., Smith, I. A.,
and Crider, A., 1997, \apjl 479, L35

\reference{linsky1} Linsky, J. L., 1990, in {\it Imaging X--ray
Astronomy}, ed. M. Elvis, (New York: Cambridge University Press) 39

\reference{lo1} Loredo, T. J., 1990, in ``Maximum Entropy and Bayesian
Methods'', ed P. F. Fougere, (Dordrecht: Kluwer Academic Publishers),
81

\reference{lo2}Loredo, T. J., 1992, in {\it Statistical Challenges in
Modern Astronomy}, ed. G. J. Babu \& E. D Fiegelson (New York:
Springer-Verlag)


\reference{lu1} Lucke, P. B. 1978, A\&A, 64, 367 \par

\reference{ma1} Matteson, J. 1978, AIAA 16th Aerospace Sciences Meeting, paper
        78-35 \par

\reference{ma2} Matz, S. M. et al. 1985, \apjl,  288, L37 \par

\reference{mcbplumet93} McBreen, B., Plunkett, S., \& Metcalfe, L., 1993,
\aap, Suppl., 97, 81 \par

\reference{me97} M\'esz\'aros, P. \& Rees, M. J., 1997,
\apj, 476, 232

\reference{metzger} Metzger, M. R., et al., 1997, Nature, 387, 878

\reference{mo1} 
Motch, C., Belloni, T., Buckley, D., Gottwald, M., Hasinger, G.,
Pakull, M.W., Pietsch, W., Reinsch, K., Remillard, R.A.,
Schmitt, J.H.M.M., Tr\"umper, J., \& Zimmerman, M.W., 1991a,
\aap, { 246}, L24 \par

\reference{mo2}Motch, C., Stella, L., Lanot-Pachero, E., Mouchet, M., 1991b,
{\it Ap.J.}, {\bf 369}, 490


\reference{mu1} Murakami, T., et al. 1988, \nat, 335, 234 \par

\reference{mu2} Murakami, T., et al. 1991, \nat, 350, 592 \par

\reference{mu3} Murakami, T., Inoue, H., van Paradijs, J., Fenimore,
E., \& Yoshida, A., 1992, in {\it Gamma--Ray Bursts}, ed. C. Ho,
R. I. Epstein \& E. Fenimore (Cambridge: Cambridge University Press),
239


\reference{mu97} Murakami, T., et al., 1998, in
{\it Proceedings of the 4th Huntsville Symposium on Gamma--ray Bursts}
AIP Conference Proc. 428, Meegan, C. A., Preece, R. D. \& Koshut, T. M.
(New York: AIP), 435 \par


\reference{no1} Norris, J.P., Share, G. H., Messina, D. C., Dennis, B. R.,
Desai, U. D., Cline, T. L., Matz, S. M., \& Chupp, E. L., 1986,
\apj, 301, 213 \par

\reference{no2} Norris, J.P., Hertz, P., Wood, K. S., \& Kouveliotou, C.
1991, \apj, 366, 240 \par

\reference{pa97} Paczy\'nski, B. \& Kouveliotou, C., 1997,
Nature, 389, 548

\reference{pe1} Pendleton, G. et al. 1996, in 
{\it Proc. of the 3rd Huntsville Symposium on Gamma--Ray Bursts
AIP Conf. Proc. 384},
ed. C. Kouveliotou, M. F. Briggs, \& G. J. Fishman,
(New York: AIP) 384 \par

\reference{pe97} Pendleton, G. et al., 1997, \apj, 489, 175

\reference{pillaloeb} Pilla, R. \& Loeb, A., 1997, in Proc. of the
VIII Marcel Grossman Meeting on General Relativity, eds. R. Ruffini
\& T. Piran (Singapore: World Scientific) in press


\reference{piro} Piro, L. et al., 1998a, in
{\it Proceedings of the 4th Huntsville Symposium on Gamma--ray Bursts}
AIP Conference Proc. 428, Meegan, C. A., Preece, R. D. \& Koshut, T. M.
(New York: AIP) \par

\reference{piro97b} Piro, L., et al., 1997b, IAU Circ. 6797 \par

\reference{piro98} Piro, L. et al. 1998b; 
{\it Bulletin of the American Astronomical Society}, 29, 1303 \par

\reference{pi95} Pizzichini, G. 1995, in Proc. 24th Int. Cosmic-Ray
Conf., Rome, Italy, 2, 81 \par

\reference{pr1} Preece, R. et al., 1996, in 
{\it Proc. of the 3rd Huntsville Symposium on Gamma--Ray Bursts
AIP Conf. Proc. 384},
ed. C. Kouveliotou, M. F. Briggs, \& G. J. Fishman,
(New York: AIP)
223 \par

\reference{pftv}Press, ~W.~H., Flannery, ~B.,~P., Teukolsky, ~S.~A.,
and Vetterling, ~W.~T., 1986, {\it Numerical Recipes} (New York:
Cambridge University Press)


\reference{ri1} Ricker, G. R. et al., 1992, in 
{\it Gamma--Ray Bursts: Observations, Analyses and Theories}
ed. C.~Ho, R.~I.~Epstein \& E.~E.~Fenimore 
(Cambridge: Cambridge University Press) 288 \par


\reference{ro1} Rothschild, R. {\it et. al.}, 1979, {\it SpScInstr}, { 4}, 269 \par

\reference{saripiran} Sari, R. \& Piran, T., 1996, MNRAS, 287, 110

\reference{scheid1}Scheid, Francis J, 1968, {\it Schaum's Outline
Series of Numerical Methods}, (New York: McGraw-Hill Book Company).


\reference{smith97} Smith, D., et al., 1998, in
{\it Proceedings of the 4th Huntsville Symposium on Gamma--ray Bursts}
AIP Conference Proc. 428, Meegan, C. A., Preece, R. D. \& Koshut, T. M.
(New York: AIP), 430 \par

\reference{st1} Strohmayer, T. E., Fenimore, E. E., Murakami, T., \&
Yoshida, A., 1998, {\apj} in press. \par

\reference{ta97} Takeshima, Y., et al., 1998, in
{\it Proceedings of the 4th Huntsville Symposium on Gamma--ray Bursts}
AIP Conference Proc. 428, Meegan, C. A., Preece, R. D. \& Koshut, T. M.
(New York: AIP), 414 \par

\reference{te1} Tennant, A.F., 1983,
{\it Rapid X--ray Variability of Active Galaxies},
Ph.D. Thesis, University of Maryland \par

\reference{te2} Terrell, J., Fenimore, E. E., Klebesadel, R. W., \& 
Desai, 1982 \apj, 254, 279 \par

\reference{van} van Paradijs, J. et al., 1997, Nature, 386, 686

\reference{vi97} Vietri, M., 1997, \apjl, 488, L105 \par

\reference{wh1} Wheaton, W. A., et al. 1973 \apjl, 185, L57 \par

\reference{yaq96} Yaqoob, T., 1997, \apj, 479, 184 \par

\reference{yo1} Yoshida, A., et al. 1989, \pasj, 41, 509 \par

\reference{yo2} Yoshida, A., Murakami, T., Nishimura, J., Kondo, I,
and Fenimore, E. E., 1992, in
{\it Gamma--Ray Bursts: Observations, Analyses and Theories}, 
ed. C. Ho, R. I. Epstein, \& E. E. Fenimore,
(Cambridge: Cambridge University Press), 399 \par

\end{references}
\end{document}